\documentclass[aps,prb,superscriptaddress,showpacs,floatfix]{revtex4-1}
\usepackage{amsfonts}
\usepackage{graphicx,amsmath,amssymb,xspace,epsfig,float,multirow,subfigure,tabularx}

\begin{document}

\title{Charge conserving approximation for excitation  properties
of crystalline materials.}
\author{Baruch Rosenstein}
\email{baruchro@hotmail.com}
\affiliation{Electrophysics Department, National Chiao Tung University, Hsinchu 30050,
\textit{Taiwan, R. O. C}}
\author{Dingping Li}
\email{lidp@pku.edu.cn}
\affiliation{School of Physics, Peking University, Beijing 100871, \textit{China}}
\affiliation{Collaborative Innovation Center of Quantum Matter, Beijing, China}
\date{\today}

\begin{abstract}
A charge conserving approximation scheme determining the excitations of
crystalline solids is proposed. Like other such approximations, it relies on
``downfolding" of the original microscopic model to a simpler electronic
model on the lattice with pairwise interactions. A systematic truncation of
the set of Dyson - Schwinger equations for correlators of the low energy
(downfolded) model of a material, supplemented by a ``covariant" calculation
of correlators lead to a converging series of approximates. The covariance
preserves all the Ward identities among correlators describing various
condensed matter probes. It is shown that the third order approximant of
this kind beyond classical and gaussian (Hartree - Fock) is precise enough
and due to several fortunate features the complexity of calculation is
surprisingly low so that a realistic material computation is feasible. Focus
here is on the electron field correlator describing the electron (hole)
excitations measured in photoemission and other probes. The scheme is tested
on several solvable benchmark models.
\end{abstract}

\maketitle


\section{Introduction}

Calculation of the band structure and response functions of crystalline
material with determined chemical composition is one of the most important
theoretical problems in condensed matter physics. However computing the
electronic excitations and spectra of a stoichiometric chemically
well-defined compounds with significant correlations from first-principles
continues to be a major challenge in computational material science.
Historically the Kohn-Sham density functional method\cite{Kohn} (DFT) opened
the door to such calculations. The basic many-body Hamiltonian is that of
the jellium model (neglecting the phonon degrees of freedom). The method
approximates the many-body physics by a noninteracting electrons is periodic
potential. It is successful to map out general features of the band
structure of numerous crystalline solids.

However, DFT is not accurate enough in the most important (for condensed
matter physics) range of energies near the Fermi level for which many-body
effects are important. Kohn-Sham eigenvalues have been used to interpret the
single particle excitation energies measured in direct and inverse
photoemission experiments. Reasonable results were obtained in simple
metals, however, when the excited state properties of semiconductors and
insulators are concerned, ambiguities between different DFT approaches (for
example the exchange correlation functional) and significant deviations from
the measured characteristics appear. A well-known example is the systematic
large (up to factor 2) underestimation of the fundamental band gaps of
semiconductors in LDA.

Therefore to zoom in on this energy range relevant for description of the
electromagnetic, thermal and other condensed matter properties that is
dominated by the excitation effects, a two step strategy is employed. DFT is
used as a first step to determine the \textquotedblleft
downfolded\textquotedblright \ model\cite{Casula}, or ``effective low energy"
electronic model containing most of the relevant information. The model is
defined on the unit cell lattice with spectrum described by the two body
electronic Greens function $G_{0}^{AB}(¦Ø ,p)$ with relevant bands
indexed by $A=1,..N_{b}$. The number of atomic orbitals (including spin) $%
N_{b}$ should not be not large with the long range interaction described
economically by an \textquotedblleft photon Greens function $W_{0}^{AB}(%
¦Ø ,p)$. The very high or low energy modes (tens of eV away from Fermi
level) are thus \textquotedblleft integrated out\textquotedblright . To be
successful the downfolding typically utilizes the maximally localized DFT
wavefunctions\cite{Vanderbilt}. The downfolded model described in more
detail below is dynamical and thus does not allow the description of the
effective low energy system by a Hamiltonian, so the Matsubara action is
employed.

The downfolded model is still very complicated and a number of numerical
(like Monte Carlo (MC) supplemented by dynamical mean field, DMF\cite%
{Kotliar}) and field theoretical methods like GW\cite{Hedin} and FLEX\cite%
{Savrasov18} were developed. The most popular analytic method to solve the
downfolded (effective low energy) model sufficient for an accurate
calculation of the condensed matter properties has been the use of the GW%
\cite{Hedin}. The GW method corrects the band gaps and other quasi-particle
properties, such as lifetimes in a wide range of weakly correlated
semiconductors and insulators. Hybertsen and Louie\cite{Hedin} showed that
applying GW approximation as a first order perturbation to the Kohn-Sham
quasiparticles of DFT (so called one shot or $G_{0}W_{0}$) provides an
accurate description of the photoemission spectra described by the electron
Greens function. However the method has well-known limitations.

First the full nonlinear set of GW equations is notoriously costly to solve.
It, in fact, has been carried out in full only for a limited number of
models like the electron gas\cite{Holm} and the results are not dramatically
better and sometimes worse than those of various GW simplifications like the
one shot $G_{0}W_{0}$. This is a set of nonlinear equations with generally a
large number of solutions. Second, it appears that results are worse in
moderately coupled materials and definitely inaccurate or even misleading in
strongly coupled electronic systems. One therefore is required either to
combine it with other methods (like Monte Carlo) or to go beyond the GW
approximation.

It has been proven difficult to systematically improve the GW approximation
by including higher-order Feynman diagrams, the so-called vertex
corrections. While extensions of the GW approach have been developed for
specific applications such as the cumulate expansion of the time dependent
Green's functions for the description of plasmon satellites\cite{Gunnarson},
or the Bethe - Salpeter approach \cite{BS}. Particularly troublesome problem
for various extensions has been to conform to the most basic principles like
charge conservation\cite{Baym}. Many approaches violate the so-called Ward
identities that are consequences of the symmetries of the system. As a
result, there exists currently no universal, viable and applicable
\textquotedblleft beyond-GW\textquotedblright \ approach\cite{Takada}.

A general method to preserve the Ward identities in an approximation scheme
was developed long time ago\cite{Kovner} in the context of field theory as
the covariant gaussian approximation (CGA) to solve an unrelated problem in
quantum field theory and superfluidity\cite{Kovner2}. A non-perturbative
variational gaussian method originated in quantum mechanics of atoms and
molecules in relativistic theories like the standard model of particle
physics had several serious related problems. First, the wave function
renormalization required a dynamical description. Second, the Green's
functions obtained using the naive gaussian approximation violated the
charge conservation. In particular the most evident problem is that the
Goldstone bosons resulted from spontaneous breaking of continuous symmetry
are massive. The method is thus considered dubious and/or inconsistent. Both
problems were solved by an observation that the solution of the minimization
equations are not necessarily equivalent to the variational Green's
function. This constitutes the covariant gaussian approximation or CGA\cite%
{Kovner}.

The method was compared with available exact results for the S-matrix in the
Gross-Neveu model\cite{GN} (a local four Fermion interactions in 1D Dirac
excitations recently considered in condensed matter physics) and with Monte
Carlo simulations in various scalar models, see ref.\onlinecite{Wang17} for
detailed description and application to thermal fluctuations in
superconductors in the framework of the Ginzburg - Landau - Wilson order
parameter approach\cite{RMP}). Applied to the electronic field correlator in
electronic systems, CGA becomes roughly equivalent to Hartree - Fock
approximation that is generally not precise enough. It's covariance might
improve the calculation of the four fermion correlators like the density-
density, but to address quantitatively photoemission or other direct
electron or hole excitation probes, a more precise method is needed.

The CGA approach is just the second in a sequence of approximations based on
covariant truncations of the DS equations in which cumulant of third and
higher order are discarded. One can continue to the next level by retaining
the third cumulant (discarding the fourth) etc. We term this approximation
covariant cubic approximation, CCA. The covariance still preserves all the
Ward identities, so it is conserving according to definition of ref. %
\onlinecite{Baym}. Up to now methodology of this kind has not been applied
to more microscopic description of realistic condensed matter systems.

The subject of the present paper is to inquire whether is possible and
computationally feasible. It is shown using several solvable benchmark
models, that the third order approximation is precise enough. Its complexity
when applied to a realistic material calculation is estimated. The focus is
on the electron (hole) excitations correlator described by the electron
field correlator that in turn can be compared to photoemission data and
other probes, although higher correlators like the density - density (or
conductivity) can be also considered as shown in refs. %
\onlinecite{Kovner,Wang17}.

The paper is organized as follows. In section II the sequence of covariant
approximations developed using the simplest possible case: the one
dimensional integral. Next in section III the third approximation of these
series, CCA is applied to a ($Z_{2}$ invariant) statistical Ginzburg -
Landau - Wilson model\cite{Lubensky} describing various statistical
mechanical systems like the Ising chain in terms of low energy (effective)
bosonic field theory. The results are compared with exact (at low dimension)
and MC simulations (higher dimensionality). In section IV the general
formalism for downfolded electronic system describing crystalline materials
is presented and applied in Section V to some low dimensional benchmark
systems like the single band Hubbard model. In Section VI contains an
estimate of complexity of application of CCA to a realistic material and
conclusions.

\section{Hierarchy of conserving truncations of DS equations}

The main ideas behind the covariant approximants are presented in this
section in the simplest possible setting. Later the third in a series of
such approximant for a many - body system will be considered in some detail.

\subsection{An exactly solvable ``bosonic" model: one dimensional integral}

To clearly present the general covariant approximation scheme, we will make
use of the simplest nontrivial model: statistical physics of a one
dimensional classical chain that is equivalent to the quantum mechanics of
the anharmonic oscillator in the next section. Our starting point here will
be the following ``free energy" as a function of a single (real) variable $%
\psi $%
\begin{equation}
f=\frac{a}{2}\psi ^{2}+\frac{b}{4}\psi ^{4}-J\psi \text{.}  \label{f}
\end{equation}%
Here $a$ and $b$ represent spectrum and ``couplings" respectively, while the``source" or ``external field" $J$ will be used to calculate correlations. The
exact partition function of just one ``fluctuating" bosonic variable is\cite%
{footnote1}

\begin{equation}
Z\left[ J\right] =\int_{\psi =-\infty }^{\infty }e^{-f}=e^{-F\left[ J\right]
}.  \label{Z[J]}
\end{equation}%
Correlators (Greens function) are defined as

\begin{equation}
G_{n}=\frac{d^{n}}{dJ^{n}}Z=\left \langle \psi ^{n}\right \rangle
\label{Gndef}
\end{equation}%
so that

\begin{eqnarray}
G_{1} &=&\frac{1}{Z}\int \psi e^{-f}=\left \langle \psi \right \rangle ,
\label{GaG2} \\
G_{2}^{c} &\equiv &G=Z^{-1}\int \psi ^{2}e^{-f}-\left \langle \psi \right
\rangle ^{2}.  \notag
\end{eqnarray}

While the odd correlators in the $Z_{2}$ symmetric case vanish, the exact
one - body correlator is

\begin{equation}
G=\frac{\sqrt{\pi }}{2Zb^{3/4}}HypergeometricU\left[ \frac{3}{4},\frac{1}{2},%
\frac{a^{2}}{4b}\right] \text{,}  \label{exactG}
\end{equation}%
where the partition function itself is%
\begin{equation}
Z=\sqrt{\frac{a}{2b}}exp\left[ \frac{a^{2}}{8b}\right] BesselK\left[ \frac{1%
}{4},\frac{a^{2}}{8b}\right] \text{.}  \label{exactZ}
\end{equation}%
The dependence of the correlator on $b$ for $a=1$, is given in Fig.1 as a
red line.

Another important set of quantities include cumulant\cite{Lubensky} defined
via the ``effective action", the Legendre transform, $A_{eff}\left( \psi
\right) =F\left[ J\right] +J\psi $, $\psi =-\frac{d}{dJ}F\left[ J\right] ,$
\ $J=\frac{d}{d\psi }A_{eff}\left[ \psi \right] $. The (two particle
irreducible) cumulants
\begin{equation}
\Gamma _{n}=\frac{d^{n}}{d\psi ^{n}}A_{eff}=\frac{d^{n-1}}{d\psi ^{n-1}}J%
\text{.}  \label{Gammandef}
\end{equation}%
The well known relations between the cumulants and correlators used below
are given in Appendix A.

\begin{figure}[tbp]
\begin{center}
\includegraphics[width=12cm]{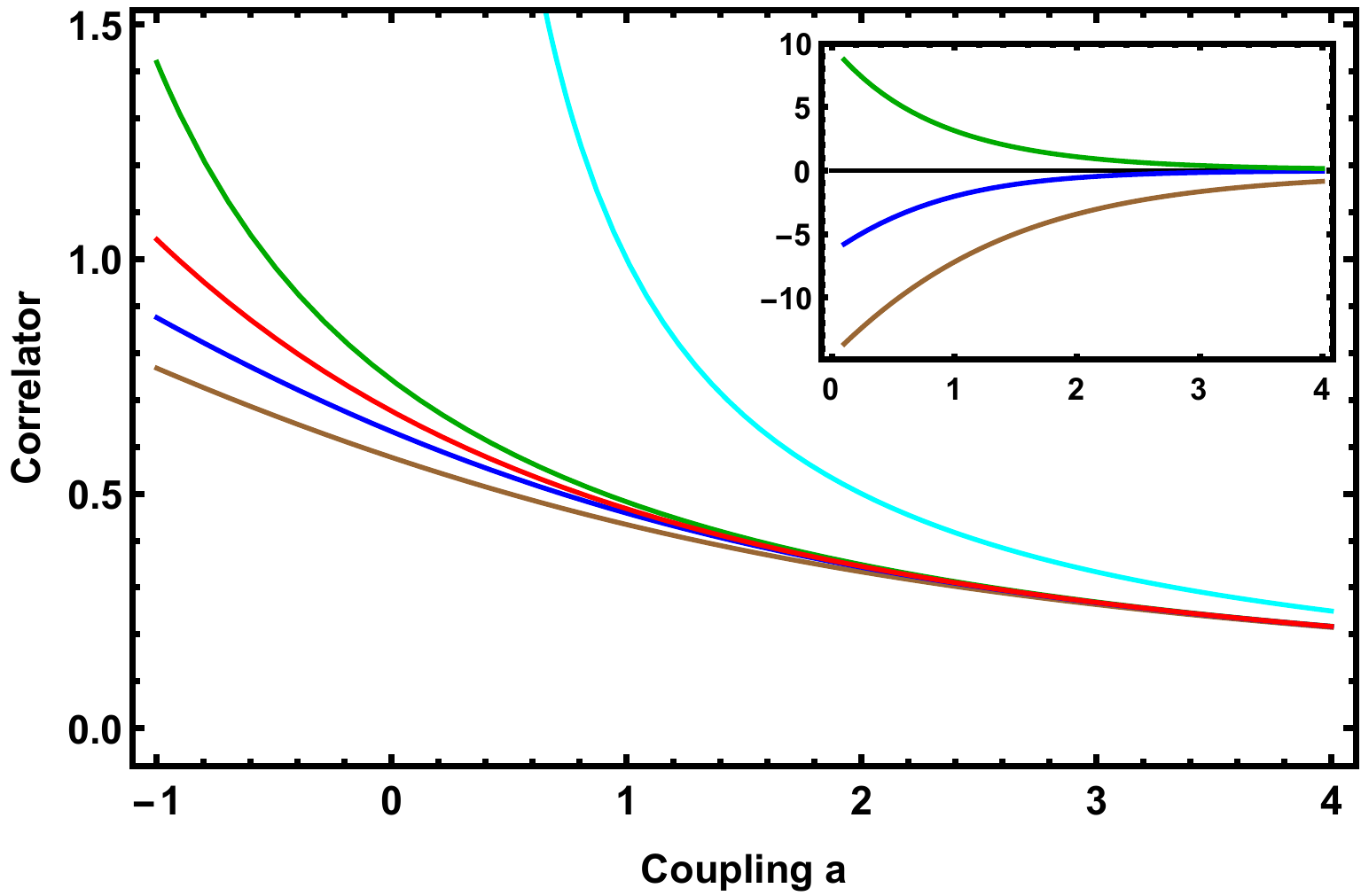}
\end{center}
\par
\vspace{-0.5cm}
\caption{Comparison of a series of successive covariant approximations for a
simple integral representing in a nutshell the $Z_{2}$ symmetric statistical
(or quantum many - body quantum) system. The red line is the exact
correlator, while the cyan, brown, green and blue are classical, gaussian,
cubic and quatic approximants respectively. Inset shows deviations (in \%)
from the exact correlator.}
\end{figure}

\subsection{The set of the DS equations}

The first in a series of the DS equations, the off shell ``equation of state"
(ES, the term ``off shell" in this paper meaning that the quantity depends on
the external source $J$) \ is

\begin{equation}
0=-\int \frac{d}{d\psi }e^{-f}\rightarrow J=\frac{1}{Z}\int \left( a\psi
+b\psi ^{3}\right) e^{-f}=a\psi +b\left \langle \psi ^{3}\right \rangle
\text{.}  \label{ESderiv}
\end{equation}%
Using the connected correlators\cite{Lubensky} (marked by subscript $c$) and
eventually cumulants, one obtains

\begin{equation}
J=a\psi +b\psi ^{3}+3b\psi G+bG_{3}^{c}\text{.}  \label{ES}
\end{equation}

Higher order DS equations in the cumulant form are obtained by
differentiating the equation above. The second DS equation is,

\begin{eqnarray}
\Gamma &=&a+3b\psi ^{2}+3bG+3b\psi \frac{d}{d\psi }G+b\frac{d}{d\psi }%
G_{3}^{c}  \label{secondDS} \\
&=&a+3b\psi ^{2}+3bG+3b\psi \Gamma G_{3}^{c}+b\Gamma G_{4}^{c}\text{,}
\notag
\end{eqnarray}%
while the next is more complicated,

\begin{equation}
\Gamma _{3}=6b\psi +6b\Gamma G_{3}^{c}-3b\psi \Gamma ^{3}G_{3}^{c2}+3b\psi
\Gamma ^{2}G_{4}^{c}-b\Gamma ^{3}G_{3}^{c}G_{4}^{c}+b\Gamma ^{2}G_{5}^{c}%
\text{.}  \label{thirdDS}
\end{equation}%
Furthermore the fourth DS (disregarding odd condensates, as they will not be
required for our purposes) has a form:

\begin{equation}
\Gamma _{4}=6b+9b\Gamma ^{2}G_{4}^{c}-b\Gamma ^{4}\left( G_{4}^{c}\right)
^{2}+b\Gamma ^{3}G_{6}^{c}\text{.}  \label{4thDS}
\end{equation}

The infinite set of DS equations is not useful in practice unless a way to
decouple higher order equations is proposed. For example, one can ignore all
the $G_{3}$ and $G_{4}$ terms in Eq.(\ref{ES}) and Eq.(\ref{secondDS}) so
that the remaining unknown variables can be solved by the two on - shell ($%
J=0$) \textquotedblleft truncated" DS equations (or equivalently the
minimization equations in gaussian variational method): the shift equation
and gap equation. This simple truncation procedure called gaussian
approximation, as stated before, is not symmetry - conserving. Fortunately a
simple improvement based on gaussian approximation, the covariant gaussian
approximation\cite{Wang17} (CGA), includes \textquotedblleft chain
corrections" to the two - body cumulant by taking functional derivative of
the off - shell (keep finite source $J$) shift equation with respect to $%
\varphi $. The chain correction is then explicitly calculated by taking
derivative of the gap equation.

In the following several subsections, a hierarchy of approximations defined
as truncations of the DS equations as well as their variational
interpretations are introduced. The CGA scheme will immediately follow when
one is familiar with the classical and gaussian approximations. Let's start
with the simplest truncation: classical approximation evaluation.

\subsection{Classical approximation}

The classical approximation consists of neglecting the two and higher body
correlators in the equation of state, Eq.(\ref{ES}),
\begin{equation}
J=a\varphi +b\varphi ^{3}\text{,}  \label{offcl}
\end{equation}%
so that the second and higher equations are decoupled from the first. Then
the \textquotedblleft minimization equation", that is just the on-shell ($%
J=0 $) ES, $a\varphi +b\varphi ^{3}=0,$is solved. For $a<0$ there are
typically several solutions of this equation \cite{footnote1} . Here
restricting discussion here to $a>0$, and the solution has $\varphi =0$.

Note that despite the fact that the minimization principle involved only the
one - body cumulant, $\varphi $, one can still calculate the higher
cumulants within the classical approximation. These are given by derivatives
of the source $J$ with respect to $\varphi $ in truncated ES, Eq.(\ref{ES}),
\begin{subequations}
\begin{equation}
\Gamma _{\left( I\right) }=\frac{\delta J}{\delta \varphi }=a+3b\varphi
^{2}=a\text{.}  \label{Gammacl}
\end{equation}%
The full correlator in momentum space is just $G_{\left( I\right) }=1/a$.
The independence on $b$ for $a=1$, is given in Fig. 1 as the cyan line,
compared to the exact correlator (red), emphasizes the fact that the
classical approximation correlator ignores the quartic term and thus might
be useful (as a starting point of the ``loop expansion", see Section IV) only
at small $b$.

The classical minimization equation can be interpreted variationally as
optimizing the free energy Eq.(\ref{f}). One can do better. Why not optimize
also the connected correlator $G$ in addition to the VEV of the field $\psi $%
? This is the gaussian approximation idea proposed early on in the context
of quantum mechanics and develop in field theory in eighties of the last
century, see ref.\onlinecite{Stevenson} and references therein.

\subsection{Covariant gaussian approximation}

Now we drop in the first two DSE all the three field cumulants (equivalently
connected correlators). This leaves us with the coupled equation for the two
variational parameters

\end{subequations}
\begin{eqnarray}
J &=&a\psi +b\psi ^{3}+3b\psi G^{tr};  \label{gaussmin} \\
\Gamma ^{tr} &=&a+3b\psi ^{2}+3bG^{tr}\text{.}  \notag
\end{eqnarray}%
The first equation is obviously obeyed for $\psi =0$, while the second takes
a form

\begin{equation}
3bG^{tr2}\ =1-aG^{tr}\rightarrow G^{tr}=\frac{-a+\sqrt{a^{2}+12b}}{6b}\text{.%
}  \label{gap}
\end{equation}%
Within the covariant approximation described in detail in ref.%
\onlinecite{Wang17}, the connected correlator $G_{\left( II\right) }\ $is
equal to $G^{tr}$. The symmetric solution exists for any $a$, however
spurious first order transition to the ``symmetry broken" solution occurs at $%
a_{sc}=-\sqrt{6b}$.

The dependance of the correlator for $b=1$ on $a$\ in the range $-1<a<4$ is
given in Fig. 1 as the brown line. It is significantly better than
classical, yet underestimates the correlator up to 15\%, see inset at $a=0$.
This value already approaches the spurious transition at $a_{sc}=-\sqrt{6b}$%
. The approximation becomes better in the perturbative region at large $a$,
as will be discussed later.

\subsection{The third order (cubic) approximation}

Continuing the same idea the neglect of fourth and higher correlators. The
ES of state is now exact,%
\begin{equation}
J=a\psi +b\psi ^{3}+3b\psi G^{tr}+bG_{3}^{tr}\text{,}  \label{ESfull}
\end{equation}%
while the next two are approximate (truncated),%
\begin{eqnarray}
\Gamma ^{tr} &=&a+3b\psi ^{2}+3bG+3b\psi \Gamma ^{tr}G_{3}^{tr}\text{;}
\label{cubicmin2} \\
\Gamma _{3}^{tr} &=&-G_{3}^{tr}\Gamma ^{tr3}=6b\psi +6b\Gamma
^{tr}G_{3}^{tr}-3b\psi \Gamma ^{tr3}G_{3}^{tr2}\text{.}  \label{cubicmin3}
\end{eqnarray}%
The first (taken on shell, $J=0$) equations are solved by $\psi =0,\Gamma
_{3}^{tr}=G_{3}^{tr}=0$. Then the gap equation coincides with the gaussian,
Eq.(\ref{gap}), with the same solution Eq(\ref{gaussmin}). However,
according to the general covariant approach outlined in ref. %
\onlinecite{Wang17}, the calculation of correlators starts with the \textit{%
off shell} ES, as in original definition in the second line of Eq.(\ref%
{Gammandef}).

For example correction to the inverse correlator is the first derivative of
Eq.(\ref{ESfull}). After making the derivative%
\begin{equation}
\Gamma _{\left( III\right) }=a+3b\psi ^{2}+3b\psi \frac{d}{d\psi }%
G^{tr}+3bG^{tr}+b\frac{d}{d\psi }G_{3}^{tr}\text{,}  \label{CCAGamma}
\end{equation}%
one substitutes the truncated quantities and their derivative \textit{on
shell}:%
\begin{equation}
\Gamma _{\left( III\right) }=a+3bG^{tr}+b\frac{d}{d\psi }G_{3}^{tr}\text{.}
\label{CCAGam1}
\end{equation}%
The first two terms, according to the gap equation, Eq.(\ref{cubicmin2}),
are inverse of the truncated propagator $G^{tr}$, so that the cumulant can
be conveniently written as

\begin{equation}
\Gamma _{\left( III\right) }=\Gamma ^{tr}+\Delta \Gamma \text{.}
\label{deltaGamdef}
\end{equation}

In the last term, so called ``chain" correction,$\frac{d}{d\psi }G_{3}^{tr}$,
is naturally obtained from the differentiation of the (off shell multiplied
by $G^{tr3}$) truncated third minimization equation, Eq.(\ref{cubicmin3}):

\begin{equation}
0=\frac{d}{d\psi }G_{3}^{tr}+6bG^{tr3}+6bG^{tr2}\frac{d}{d\psi }G_{3}\text{.}
\label{chaineq}
\end{equation}%
Here the general relation between cumulant and connected functions, $%
G^{3}\Gamma _{3}=G_{3}\,$was used. Unlike the gap equation, this equation is
linear, so that,
\begin{equation}
\frac{d}{d\psi }G_{3}=-\frac{6bG^{tr3}}{1+6bG^{tr2}}\text{,}
\label{chaincub}
\end{equation}%
and finally%
\begin{equation}
\Delta \Gamma =-\frac{6bG^{tr3}}{1+6bG^{tr2}}\text{.}  \label{delgamcub}
\end{equation}%
\

Now the spurious second order transition to a ``symmetry broken" solution
occurs at a lower negative value $a_{sc}=-\sqrt{12b}$ than the CGA one. This
is a trend. Higher approximation symmetric phase solution works in the
increasingly large portion of the parameter space. The dependance of the
correlator for $b=1$ on $a$\ in the range $-1<a<4$ is given in Fig. 1 as the
green line. CCA now overestimates the correlator up to 10\% at $a=0$, see
inset.

\subsection{The fourth order (quartic) approximation}

The truncation is not needed now for the first two DSE, so that the ES stays
as in Eq.(\ref{ESfull}) and the gap equation takes the full form

\begin{equation}
\Gamma ^{tr}=a+3b\psi ^{2}+3bG+3b\psi \Gamma ^{tr}G_{3}^{tr}+3b\Gamma
^{tr2}G_{3}^{tr2}+b\Gamma ^{tr}G_{4}^{tr}\text{,}  \label{firstq}
\end{equation}%
while the next two are approximate. The third will be required off shell,%
\begin{equation}
\Gamma _{3}^{tr}=6b\psi +6b\Gamma ^{tr}G_{3}^{tr}-3b\psi \Gamma
^{tr3}G_{3}^{tr2}+3b\psi \Gamma ^{tr2}G_{4}^{tr}-b\Gamma
^{tr3}G_{3}^{tr}G_{4}^{tr}\text{.}  \label{secondq}
\end{equation}%
The last term is needed only on - shell, thus all the odd correlators can be
omitted:
\begin{equation}
\Gamma _{4}^{tr}=6b+9b\Gamma ^{tr2}G_{4}^{tr}-b\Gamma ^{tr4}\left(
G_{4}^{tr}\right) ^{2}\text{.}  \label{thirdq}
\end{equation}

The first and the third minimization equations are still trivially satisfied
as long as odd correlators vanish. The \ second and the fourth equations on
- shell for the two even connected correlators $G^{tr},G_{4}^{tr}$, take
(upon multiplication by $G^{tr}$ and $G^{tr4}$ respectively) the ``Bethe -
Salpeter" form%
\begin{eqnarray}
1 &=&aG^{tr}+3bG^{tr2}+bG_{4}^{tr};  \label{BS1} \\
G_{4}^{tr} &=&-6bG^{tr4}-9bG^{tr2}G_{4}^{tr}+b\left( G_{4}^{tr}\right) ^{2}%
\text{.}  \label{BS2}
\end{eqnarray}%
The gap equations, solved for $G_{4}^{tr}$ allows to obtain a cubic equation,

\begin{equation}
30b^{2}G^{tr3}+15abG^{tr2}-\left( 12b-a^{2}\right) G^{tr}-a=0\text{,}
\label{BSeq}
\end{equation}%
for $G^{tr}$.

The cumulant $\Gamma $, given by the derivative of the ES in terms of the
chain $\frac{d}{d\psi }G_{3}^{tr}$ is the same as for the cubic
approximation, Eq.(\ref{chaincub}). However the chain equation, although
still linear,%
\begin{equation}
-\frac{d}{d\psi }G_{3}^{tr}=6bG^{tr3}+6bG^{tr2}\frac{d}{d\psi }%
G_{3}^{tr}+3bG^{tr}G_{4}^{tr}-bG_{4}^{tr}\frac{d}{d\psi }G_{3}^{tr}\text{,}
\label{chainq}
\end{equation}%
now gives
\begin{equation}
\frac{d}{d\psi }G_{3}^{tr}=3G^{tr}\frac{2bG^{tr2}+bG_{4}^{tr}}{%
bG_{4}^{tr}-1-6bG^{tr2}}=3\frac{bG^{tr2}+aG^{tr}-1}{a+9bG^{tr}}\text{.}
\label{chainqsol}
\end{equation}%
The cumulant now takes a form

\begin{equation}
\Gamma _{\left( IV\right) }=a+3bG^{tr}+3b\frac{bG^{tr2}+aG^{tr}-1}{a+9bG^{tr}%
}\text{.}  \label{quartres}
\end{equation}%
Its inverse for $b=1$ is given in Fig. 1 as the blue line over the range $%
-1<a<4$. It underestimates the exact results by just 5\% at $\ a=0$, as
shown in the inset. The general trend is that the approximants oscillate
converging the exact result. Let us now discuss the convergence of these
approximations to the exact correlator, Eq.(\ref{exactG}) and their
asymptotic at weak and strong coupling.

\section{Testing the covariant approximations on statistical physics models}

In this section the results of the covariant approximants outlined above are
compared with exact values (or in more complicated cases numerical
simulations that is known to be reliable) for the bosonic $Z_{2}$ invariant
Ginzburg - Landau -Wilson models. The formalism is generalized to the
lattice model of arbitrary dimension $D$. We start with $D=0$.

\subsection{Convergence of the first four approximants to the exact
correlator of the bosonic toy model.}

The ``partition function" of the toy model, used in the previous section, Eq.(%
\ref{Z[J]}), despite having two coefficients, $a$ and $b$, has just one
independent parameter: $a/\sqrt{b}$. Since $b>0$ and we first assume $a>0$,
only $a=1$ is considered in Fig.1. The figure indicates that the sequence of
approximants converges quite fast. To make this more quantitative, let us
first compare asymptotic.
\begin{table*}[tph]
\caption{Weak and strong coupling expansions of covariant approximations for
the toy model.}
\begin{center}
\begin{tabular}{|l|l|l|l|l|l|}
\hline
approximants & weak coupling expansion & strong coupling expansion & b=1 &
b=4 & b=16 \\ \hline
exact & $1-3b+24b^{2}-297b^{3}+4896b^{4}$ & $0.67598b^{-1/2}-0.27153b^{-1}$
&  &  &  \\ \hline
classical \  \  \  \ (I) & $1+0$ $b+0$ $b^{2}+0$ $b^{3}+0$ $b^{4}$ & $1$ & 114
& 259 & 559 \\ \hline
cov. gauss \ (II) & $1-3b+18b^{2}-135b^{3}+1134b^{4}$ & $0.57735b^{-1/2}$ $-$
$0.16667b^{-1}$ & -7.2 & -10.3 & -12.3 \\ \hline
cov. cubic \ (III) & $1-3b+24b^{2}-261b^{3}+3222b^{4}$ & $0.74231b^{-1/2}$ $%
-0.37755b^{-1}$ & 3.1 & 5.5 & 7.3 \\ \hline
cov. quartic (IV) & $1-3b+24b^{2}-297b^{3}+4536b^{4}$ & $0.63246b^{-1/2}$ $%
-0.20833b^{-1}$ & -2.0 & -3.7 & -4.9 \\ \hline
\end{tabular}%
\end{center}
\end{table*}

At small coupling $b<<a^{2}$, the expansion up to $b^{4}$ are given in Table
I. One observes that the expansion is exact to order $n^{N-1}$, where $N$ is
the order of the approximation. A more surprising result is that the leading
incorrect term is within $10\%$ of the correct value. For example for the
cubic approximation the $b^{3}$ coefficient is $261$ compared with exact $%
297 $, the quartic approximation the $b^{4}$ coefficient is $4536$ compared
with exact $4896$.

At strong coupling the situation is a bit different. At any order of the
expansion parameter $1/\sqrt{b}$ the coefficient converges to the exact at
large $N$. In between (see values $b=1,4$ and $16$ in Table I).\ The
deviation from exact (given in \%) does not exceed 13\% for covariant
gaussian (typical of course to numerous ``mean field" approaches), 8\% for
covariant cubic and 5\% for covariant quartic. Note that the deviations
would increase dramatically if a noncovariant (naive or variational) version
is used\cite{Wang17}.

To conclude, increasing the rank of the covariant truncation approach
increases precision at a price of more complexity. Now we consider the same $%
\phi ^{4}$ Ginzburg - Landau - Wilson (GLW) model in higher dimensions $D>0$%
. Although an exact correlator is unknown, it can be calculated numerically
with practically arbitrary precision as in ref. \onlinecite{Wang17} and
compared with classical£¬ CGA and CCA approximations.

\subsection{Statistical mechanics of the D - dimensional $\protect \psi ^{4}$
model}

The statistical physics in terms of the (real) order parameter of the Ising
universality class\cite{Lubensky,Kleinert,Rothe} is defined by the
statistical sum $Z=\exp \left[ -A/T\right] $ on a hypercube lattice $%
r=\left
\{ r_{1},...r_{D}\right \} $ $r_{i}=1,...N$:

\begin{equation}
A=d^{D}\left \{ -\frac{1}{2}\sum \nolimits_{r,r^{\prime }}\psi _{r}\nabla
_{r-r^{\prime }}^{2}\psi _{r^{\prime }}+\sum \nolimits_{r}\left( \frac{a}{2}%
\psi _{r}^{2}+\frac{b}{4}\psi _{r}^{4}-J_{r}\psi _{r}\right) \right \} .
\label{MCaction}
\end{equation}%
The lattice laplacian in $D$ dimensions is%
\begin{equation}
\bigtriangledown _{r}^{2}=d^{-2}\sum \nolimits_{i=1,...,D}\left( \delta _{r-%
\widehat{i}}+\delta _{r+\widehat{i}}-2\delta _{r}\right) \text{.}
\label{latLaplacian}
\end{equation}%
The hopping direction is denoted\cite{Rothe,Kleinert} by $\widehat{i}$.
Periodicity of each direction is assumed, with ``lattice spacing" setting the
length scale $d=1$.

The temperature will set the energy scale $T=1$. This can be regarded as a
lattice version of the Ginzburg - Landau - Wilson action\cite%
{Lubensky,Kleinert} and is a generalization of the toy model of the previous
subsection, in that the position index $r$ appears. In fact we transform it
to the momentum space%
\begin{equation}
\psi _{r}=\frac{1}{N^{D/2}}\sum_{k_{i}=1}^{N}\exp \left[ \frac{2\pi i}{N}%
k\cdot r\right] \psi _{k}\text{,}  \label{Fourierdef}
\end{equation}%
so that the

\begin{eqnarray}
A &=&\frac{1}{2}\sum \nolimits_{k}\varepsilon _{k}\psi _{k}\psi _{-k}+\frac{b%
}{4V}\sum \nolimits_{k_{1}k_{2}k_{3}}\psi _{k_{1}}\psi _{k_{2}}\psi
_{k_{3}}\psi _{-k_{1}-k_{2}-k_{3}};  \label{momentumspace} \\
\varepsilon _{k} &=&\widehat{k}^{2}+a;\text{ \  \  \ }\widehat{k}^{2}=4\sum
\nolimits_{i=1}^{D}\sin ^{2}\left[ \frac{\pi k_{i}}{N}\right] \text{.}
\notag
\end{eqnarray}%
where $V\equiv N^{D}$ is the number of points of the lattice.

The off shell truncated DS equations now take a form:

\begin{eqnarray}
J_{-p} &=&\varepsilon _{p}\psi _{-p}+\frac{b}{V}\left( \psi _{k_{1}}\psi
_{k_{2}}\psi _{-k_{1}-k_{2}-p}+3G_{k_{1}k_{2}}\psi
_{-k_{1}-k_{2}-p}+G_{k_{1}k_{2},-k_{1}-k_{2}-p}\right) ;  \label{DS1D} \\
\Gamma _{qp} &=&\varepsilon _{p}\delta _{q+p}+\frac{b}{V}\left( 3\psi
_{k_{1}}\psi _{-k_{1}-q-p}+3G_{k_{1},p+q-k_{1}}+3\Gamma
_{qk_{1}}G_{k_{1}k_{2}k_{3}}\psi _{-k_{2}-k_{3}-p}\right) ;  \notag \\
\Gamma _{qpu} &=&\frac{3b}{V}\left( 2\psi _{-q-p-u}+\Gamma
_{qk_{2}}G_{k_{2},k_{1},-k_{1}-p-u}+\Gamma
_{uk_{2}}G_{k_{2},k_{1},-p-q-k_{1}}+\Gamma _{qk_{2}u}G_{k_{2}k_{1}k_{3}}\psi
_{-k_{2}-k_{3}-p}\right) \text{.}  \notag
\end{eqnarray}%
Here summations over $k_{1},k_{2},k_{3}$ should understood as an Einstein
summation index are assumed. Minimization equations in the symmetry unbroken
phase $\psi =\Gamma _{qpu}=0$ phase reduces to the gaussian one, solved in
ref.\onlinecite{Wang17}.

The $Z_{2}$ symmetric solution of the minimization equations reduce, as in
the toy model, to the solution of the gap equation, that due to translation
invariance%
\begin{equation}
\Gamma _{qp}^{tr}=\delta _{q+p}\gamma _{p};\text{ }\gamma _{p}=\widehat{p}%
^{2}+m^{2};  \label{trinvarscalar}
\end{equation}%
is algebraic%
\begin{equation}
m^{2}=a+\frac{3}{V}\sum \nolimits_{k}g_{p}.  \label{geqsc}
\end{equation}%
There $g_{p}=\gamma _{p}^{-1}$.\ Correction to correlator subsequently is%
\begin{equation}
\Delta \gamma _{p}=\frac{b}{V}\sum%
\nolimits_{k_{1}k_{2}}C_{-p;k_{1},k_{2},-p-k_{1}-k_{2}}=\frac{b}{V}\sum
\nolimits_{k_{1}k_{2}}c_{p;k_{1},k_{2}}\text{,}  \label{delgamscalar}
\end{equation}%
where the chain is defined as $C_{w;l_{1}l_{2}l_{3}}\equiv \frac{\delta }{%
\delta \psi _{w}}G_{l_{1}l_{2}l_{3}}$. The translation invariance of the
chain, $C_{w;l_{1}l_{2}l_{3}}=\delta _{l_{1}+l_{2}+l_{3}-w}c_{w;l_{1}l_{2}}$%
, allows to write the second equality.

Let us turn now to the chain equations, obtained, as in the toy model, from
(functional) derivative of the third DSE, Eq.(\ref{DS1D}). It reads

\begin{equation}
c_{w;lm}=-\frac{3b}{V}\left( 2g_{l}g_{m}g_{-l-m+w}+g_{m}g_{-l-m+w}\sum
\nolimits_{k}c_{w;lk}+g_{l}g_{m}\sum \nolimits_{k}c_{w;-l-m+w,k}\right)
\text{.}  \label{chscalar}
\end{equation}%
One notices that, due to locality of the interaction, in addition to the
fact that $w$ is a ``spectator", on the RHS of the equation only sum over the
last momentum $k$ in $c$ appears. Since we need only summed $c$ in the
correction to the inverse correlator, Eq.(\ref{delgamscalar}), we sum up the
equation over the last index $m$, $c_{w;l}\equiv \sum \nolimits_{m}c_{w;lm}$%
. Using the fish integral $f_{w}=\frac{1}{V}\sum \nolimits_{k}g_{k}g_{w-k}$,
the chain equation finally takes a form:

\begin{equation}
c_{w;l}=-3b\left( 2g_{l}f_{w-l}+f_{w-l}c_{w;l}+\frac{g_{l}}{V}\sum
\nolimits_{k}g_{k}c_{w;-l-k+w}\right) \text{.}  \label{chainscalar}
\end{equation}%
This set of $V$ linear equations is solved numerically. Let us start with a
simpler 1D case that allows an simpler solution (and can be interpreted as
quantum mechanics of the anharmonic oscillator).

\subsection{D=1 chain (or quantum mechanical anharmonic oscillator)}

The $D=1$ case corresponds to the GLW type description of the Ising chain.
It is equivalent to the quantum mechanics of the anharmonic oscillator for
small $d$ limit, see Eq.(\ref{MCaction}). This case, although not solvable
analytically, allows numerical solution with unlimited precision, and is
compared with covariant gaussian and cubic approximations first. A more
general case of arbitrary $d$ is compared with Monte Carlo simulations.

\begin{figure}[tbp]
\begin{center}
\includegraphics[width=18cm]{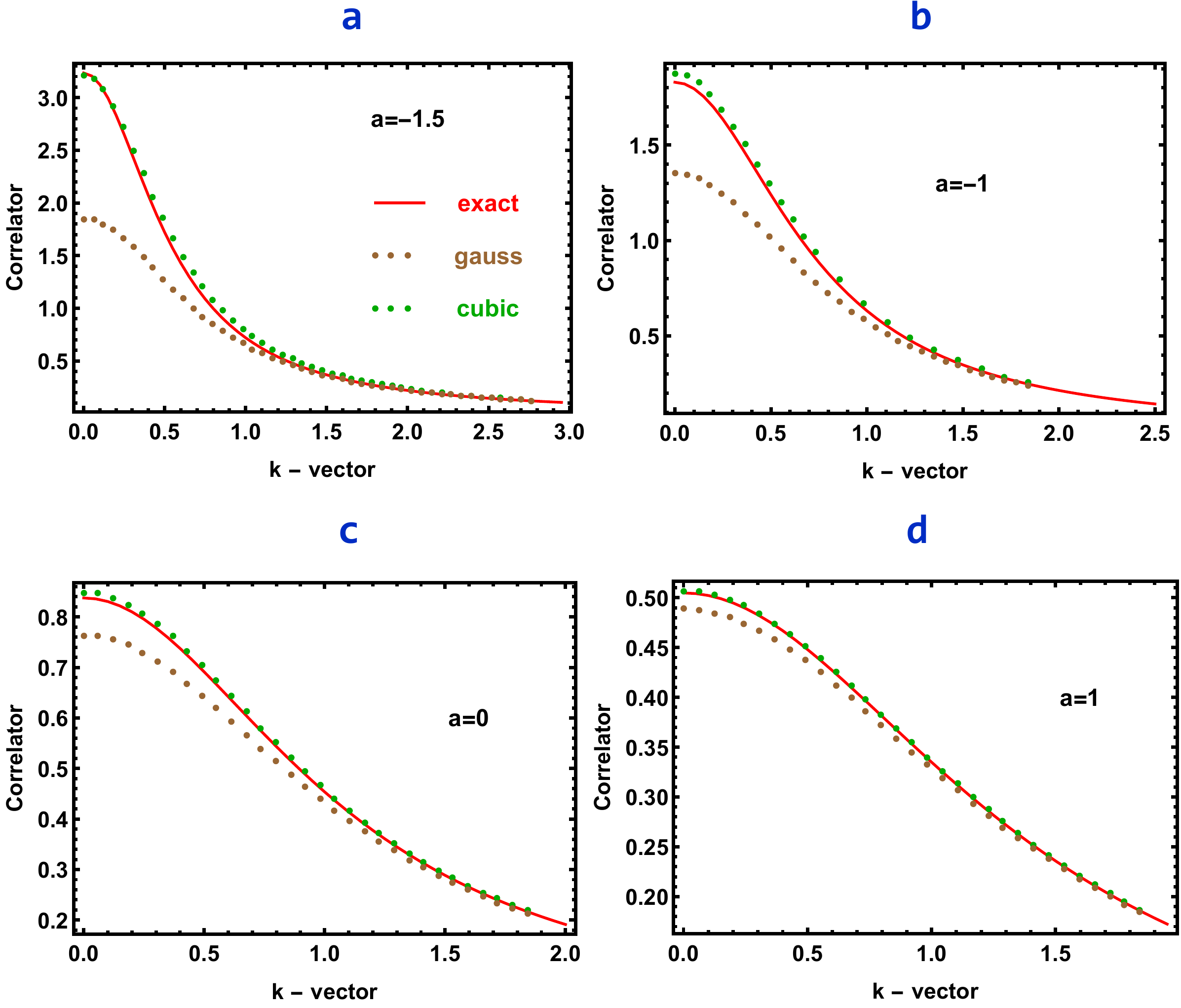}
\end{center}
\par
\vspace{-0.5cm}
\caption{Comparison of a gaussian and cubic covariant approximations for a
1D Ginzburg - Landau - Wilson chain (representing the $Z_{2}$ symmetric
statistical physics) for small $d=1/20$ with quantum mechanical anharmonic
oscillator. The red line is the exact correlator, while the brown and the
green dots are CGA and CCA respectively.}
\end{figure}

\subsubsection{Low temperature compared with the quantum anharmonic
oscillator}

In 1D the temperature $T$ in statistical physics of classical chain can be
reinterpreted as $\hbar $ and the quantum anharmonic oscillator,%
\begin{equation}
H=-\frac{1}{2}\frac{d^{2}}{dx^{2}}+\frac{a}{2}x^{2}+\frac{b}{4}x^{4}\text{,}
\label{QMHamiltonian}
\end{equation}%
discretized partition function (at temperature $T=1/N$. For very large $N$
the correlator approaches the correlation of $x\left( \tau \right) $ for the
ground state of the quantum anharmonic oscillator\cite{Kleinert}. The
thermal\ fluctuations in this interpretation are replaced by the quantum
ones. Classical approximation and CGA for this model for the correlator and
some composite correlators was worked out in ref. \onlinecite{Wang17}.

The distance between spatial points should be a small as possible, so we
take $d=1/20$. The coupling $b$ is fixed at $b=1$ (can be rescaled to this
value), while $a=$ $-1.5,-1,0,1$. The ``exact" correlator (the red line in
Fig. 2) was calculated as%
\begin{equation}
G_{k}=\left \vert \left \langle 0\left \vert x\right \vert 0\right \rangle
\right \vert ^{2}2\pi \delta _{k}+\sum \nolimits_{n>0}\left \vert \left
\langle 0\left \vert x\right \vert n\right \rangle \right \vert ^{2}\frac{%
2\left( E_{n}-E_{0}\right) }{k^{2}+\left( E_{n}-E_{0}\right) ^{2}}\text{,}
\label{QMcorr}
\end{equation}%
$E_{n}$ and $\left \vert n\right \rangle $ are eigenvalues and eigenstates
of Hamiltonian of Eq.(\ref{QMHamiltonian}). We used $N=2048$ to ensure
continuum limit. One observes that the convergence to exact value is
generally faster than in $D=0$, see Fig.1. Cubic overestimates much less
than the gaussian underestimates the correlator for all k - vectors. For
example the $a=1$,$\, \ $Fig. 2d, CCA is within 1\% for the whole range of k
- vectors. Even for negative values of $a$ CCA is very precise away from the
spurious phase transition of the gaussian approximation at $s_{spt}=-1.97$.
In was shown in ref. \onlinecite{Wang17} that the instanton calculus is
effective only for $a<-3$, so that the approximations work in the region
where no other simple approximation scheme exists. For the value of $d$ that
is not small, one cannot rely on continuum limit quantum mechanics, so the
Monte Carlo approximate method is employed.

\begin{figure}[tbp]
\begin{center}
\includegraphics[width=12cm]{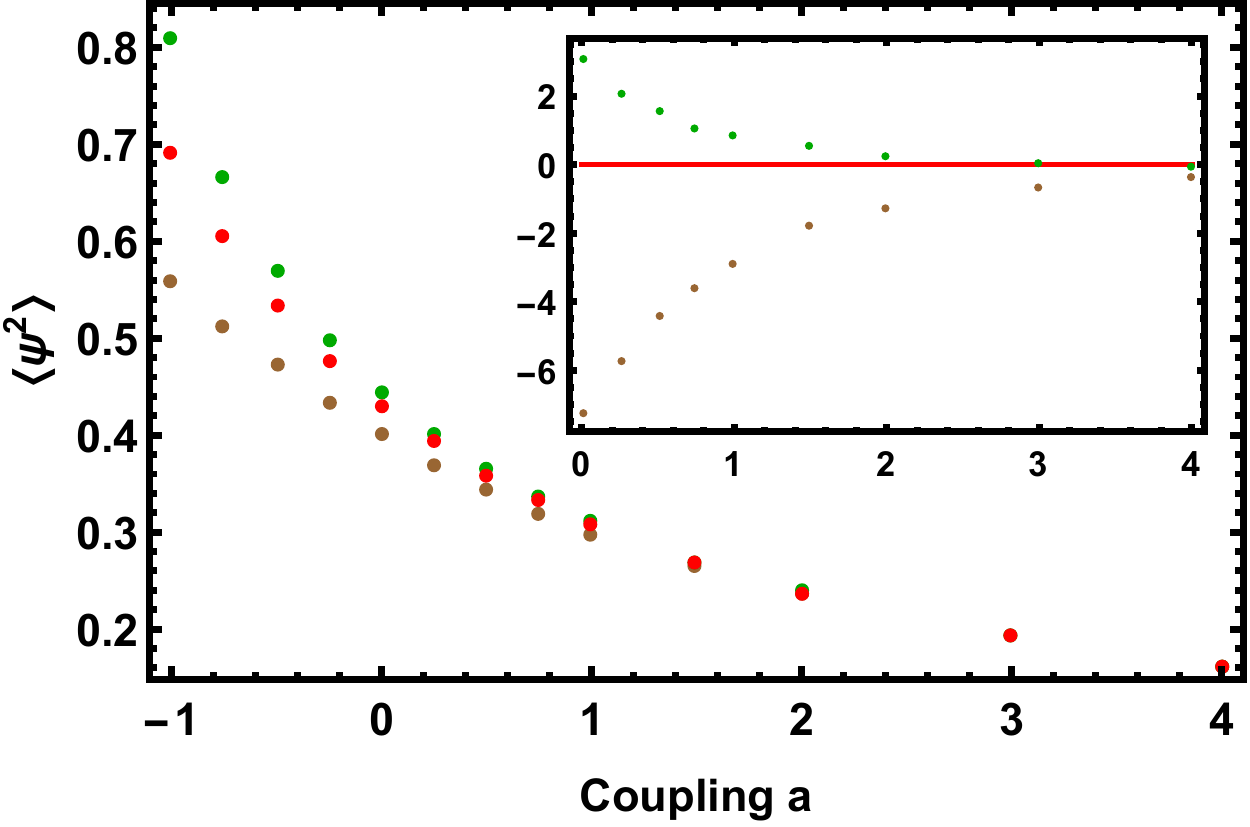}
\end{center}
\par
\vspace{-0.5cm}
\caption{Comparison of a gaussian and cubic covariant approximations for a
1D Ginzburg - Landau - Wilson chain at $d=1$. The red line is the MC
simulation result, while the brown, and the green lines are covariant
gaussian and cubic approximants respectively. Inset shows deviations (in \%)
from the exact.}
\end{figure}

\subsubsection{Monte Carlo simulation of the GLW action on finite chain}

In Fig.3 the results of the MC calculation of the average correlator in
space $\left \langle \psi _{x}^{2}\right \rangle $ are shown for $a$ in the
range $-1$ to $4$ and $b=1$. The sample size was $N=256$ with $d=1$ (with
periodic boundary condition). The standard Metropolis algorithm is usually
inefficient for $a<0$ because of the large autocorrelation of the samples.
The autocorrelation however can be reduced to a large extent by combining
the Metropolis algorithm with the cluster algorithm. This is done by using
Wolff's single-cluster flipping method\cite{Brower}. Each cycle of the Monte
Carlo iteration contains a single cluster update of the embedded Ising
variables, followed by a sweep of local updates of the original fields using
Metropolis algorithm. The calculated integrated autocorrelation time was
typically less then $10$ second in a usual desktop PC. With such reduced
autocorrelation, the statistical error for a run containing several $10^{5}$
cycles after reaching equilibrium is already small enough.

The CCA was computed for the same sample size using Mathematica (green dots
in Fig.3. The results are reminiscent of the quantum mechanical continuum
limit with maximal deviations at $a=-1$ of 6\% for CGA (underestimate)
reduced to 2\% for CCA (overestimate). A naive expectation is that, when
dimensionality is increased or interaction that becomes longer range, the
mean field - like approximations of the type considered here, the range of
applicability grows. Although in the present paper nonlocal interactions
(most notably Coulomb interactions in insulators, semiconductors) is not
considered here, the GLW model have been studied by MC in higher dimensions
(D=2,3)\cite{Arnold} and it will be compared with the CCA calculation below.

\subsection{CCA for the D=2 GLW model compared to Monte Carlo simulation}

Similar calculation has been performed in 2D for the sample size $32\times
32 $. Here in the same region of parameter space, $-1<a<4$, $b=1$, the
fluctuations influence is less pronounced, so the cluster method is not
required in the case not being too near critical state. This is above the
second order phase transition (lower critical dimensionality for the $Z_{2}$
spontaneous symmetry breaking is $D=2$) at $a_{c}=-1.1$ deduced from the
correlator $\left \langle \psi _{r^{\prime }}\psi _{r+r^{\prime
}}\right
\rangle $ as function of distance between the two points. The
correlator was averaged over 128 points $r^{\prime }$. Results for $\left
\langle \psi _{r}^{2}\right \rangle $ are presented (as the red dots) in
Fig. 4. The precision estimate for $\left \langle \psi _{r}^{2}\right
\rangle $ is 0.2\%. Thermalization was achieved after $10^{5}$ MC steps and $%
3\cdot 10^{5} $ were used for measurement.

Gaussian approximation for $\left \langle \psi _{r}^{2}\right \rangle $ was
calculated on the same lattice, see brown dots in Fig. 4. As was noticed
long ago\cite{Stevenson}, the transition at $a_{c}=-1.198$ is a spurious
weakly first order with finite excitation mass $m^{2}=0.12$ on the symmetric
side (symmetric solution exists for any $a$). This fact was one of the
problems of the approximation at the early stages of its development. The
spurious first transition however is very close to the second order
transition point found in MC. CGA underestimates the MC result by 2.5\% at $%
a=0\,$, see inset.

\begin{figure}[tbp]
\begin{center}
\includegraphics[width=12cm]{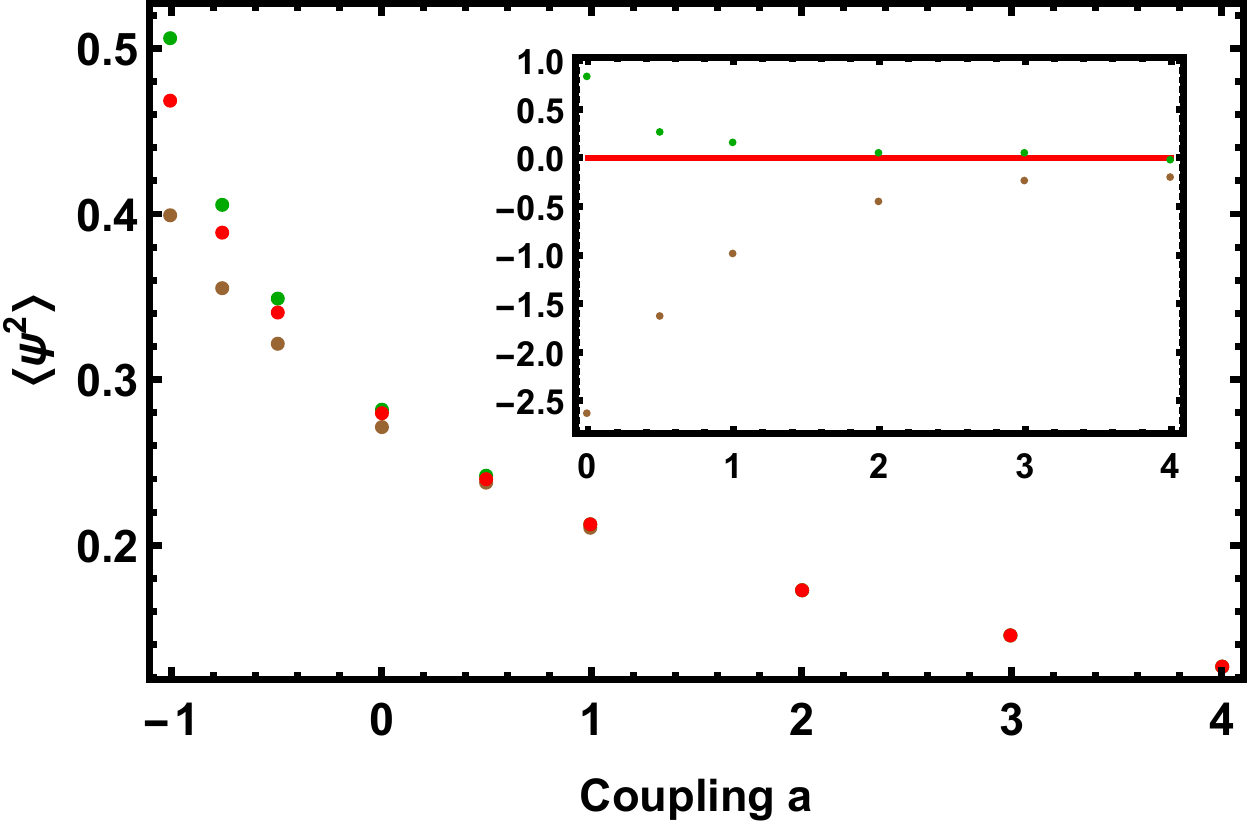}
\end{center}
\par
\vspace{-0.5cm}
\caption{The order parameter square of 2D Ginzburg - Landau - Wilson model.
MC simulation results (the red dots) are compared with covariant gaussian
(the brown dots) and cubic (green dots) approximations.}
\end{figure}

The CCA value for $\left \langle \psi _{r}^{2}\right \rangle $ in the same
range was computed for the same sample size using Mathematica (green dots in
Fig. 4) using parallel computing. The results, green dots in Fig.4,
overestimate the MC value by $0.8\%$ at $a=0$. Of course in the perturbative
region (large $a$), as before the gaussian approximation is one loop exact,
while the cubic is two - loop exact. Generally 2D convergence is better than
in 1D and is expected to further improve in 3D.

In the next section we formulate the CCA for a general fermionic model and
apply it to develop a calculational scheme for a general computation of the
electron Green function for an arbitrary crystalline material.

\section{Covariant cubic approximation for the electrons in a crystalline
solid}

Covariant gaussian approximation for a fermionic system interacting via
local four - Fermi term has been considered long time ago and compare well%
\cite{GN} with exact scattering matrix found by the factorization methods%
\cite{Zamolodchikov} in some 1+1 dimensional relativistic models (the Gross
- Neveu model, known in condensed matter physics as the Schrieffer-Su-Heeger
model, was considered). Here we formulate the third order covariant
approximation, CGA, that surprisingly turns out to be not much more
complicated computationally. The additional effort is to solve large systems
of linear equations.

\subsection{Cubic approximation in general four - fermion interaction model.}

Let us start with a rather general case using abstract notations, to
demonstrate the general structure of the method. All the characteristics of
fermionic degrees of freedom (electrons) like location in space, time,
charge, band index (including spin) etc described by (real) grassmanian
numbers are lumped into one index $A$. The four - Fermi interaction model
(described in more detail for ``downfolded" models electrons on lattice with
effective interactions in the next subsection) is defined by the Nambu -
Matsubara ``action" and ``statistical sum":
\begin{eqnarray}
\mathcal{A}\left[ \psi \right] &=&\frac{1}{2}\psi ^{A}T^{AB}\psi ^{B}+\frac{1%
}{4!}V^{ABCD}\psi ^{A}\psi ^{B}\psi ^{C}\psi ^{D};  \label{fermionaction} \\
Z\left[ J\right] &=&\int D\psi \exp \left[ -\mathcal{A}\left[ \psi \right]
+J^{A}\psi ^{A}\right] \text{.}  \notag
\end{eqnarray}

This formalism is slightly more general than the complex grassmanian numbers
approach\cite{NO}, in which a conjugate pair $\psi ,\psi ^{\ast }$ is
describing annihilation or creation of a charged fermion. The Nambu approach
is often used in description of superconducting state has an advantage of
transparency due to explicit antisymmetry of all the grassmanians. In
particular the ``hopping amplitudes" $T^{AB}$ and the interaction $V^{ABCD}$
are totally antisymmetric in generalized indices.

Let us adjust the definitions of correlators to the fermionic case, paying
attention to order of the grassmanian variables and their derivatives.
Cumulants and connected correlators of fermions are defined as

\begin{eqnarray}
\Gamma ^{A_{1}A_{2}...A_{n}} &=&\frac{\delta ^{n}\mathcal{A}^{eff}}{\delta
\psi ^{A_{1}}\delta \psi ^{A_{2}}...\delta \psi ^{A_{n}}};
\label{cumulantdef} \\
G_{c}^{A_{1}A_{2},,,A_{n}} &=&-\frac{\delta ^{n}F}{\delta J^{A_{1}}...\delta
J^{A_{n}}}=\left \langle \psi ^{A_{1}}...\psi ^{A_{n}}\right \rangle
_{c}\equiv \left \langle A_{1}...A_{n}\right \rangle \text{.}  \notag
\end{eqnarray}%
The description of CCA closely follows the steps described for bosons above.
The first is ``truncation" of the infinite set of Dyson-Schwinger equations.

\subsubsection{First three DS equations and their truncation}

Differentiating the effective action (Legendre transform of $F\left[ J\right]
\equiv -\log \left[ Z\left[ J\right] \right] $) off shell (namely in the
presence of the fermionic source $J$), the equation of motion is%
\begin{equation}
J^{A}=-\frac{\delta A^{eff}}{\delta \psi _{A}}=-T^{AX}\psi ^{X}-\frac{1}{3!}%
V^{AX_{2}X_{3}X_{4}}\left \{ \psi ^{X_{2}}\psi ^{X_{3}}\psi ^{X_{4}}+3\psi
^{X_{2}}\left \langle X_{3}X_{4}\right \rangle +\left \langle
X_{2}X_{3}X_{4}\right \rangle \right \} \text{.}  \label{ESfermion}
\end{equation}%
Note that the antisymmetry of the coefficients in the Nambu real grassmanian
used here greatly simplifies the expressions compared to the complex
grassmanian formalism. Similarly the second DS, using repeatedly the
relation,$\frac{\delta }{\delta \psi ^{B}}\left \langle
X_{1}...\right
\rangle =\Gamma ^{YB}\left \langle YX_{1}...\right \rangle $%
, is%
\begin{equation}
\Gamma ^{AB}=\frac{\delta }{\delta \psi ^{B}}J^{A}\simeq -T^{AB}-\frac{1}{2}%
\left \{ V^{ABX_{3}X_{4}}\psi ^{X_{3}}\psi
^{X_{4}}+V^{ABX_{3}X_{4}}G^{X_{3}X_{4}}-V^{AX_{2}X_{3}X_{4}}\psi
^{X_{2}}\Gamma ^{X_{1}B}\left \langle X_{1}X_{3}X_{4}\right \rangle \right
\} \text{.}  \label{secondDSferm}
\end{equation}

As in the bosonic CCA of the previous section, the fourth correlator term
was dropped from the truncated equation (this is the meaning of "$\simeq "$%
). Similarly all the terms containing fourth and fifth cumulants will be
dropped from the third DS equation:%
\begin{equation}
\Gamma ^{CAB}=\frac{\delta }{\delta \psi ^{C}}\Gamma ^{AB}\simeq -\frac{1}{2}%
\left \{
\begin{array}{c}
2V^{ABCX}\psi ^{X}+V^{ABX_{3}X_{4}}\Gamma ^{X_{1}C}\left \langle
X_{1}X_{3}X_{4}\right \rangle \\
-V^{ACX_{3}X_{4}}\Gamma ^{X_{1}B}\left \langle X_{1}X_{3}X_{4}\right \rangle
+V^{AX_{2}X_{3}X_{4}}\psi ^{X_{2}}\Gamma ^{CX_{1}B}\left \langle
X_{1}X_{3}X_{4}\right \rangle%
\end{array}%
\right \} \text{.}  \label{thirdferm}
\end{equation}%
These equations will be used twice. First the on - shell version, $J=0$, the
minimization equations are solved and then, the (CCA) correlator is computed
from a derivative of the ES using via chain rule.

\subsubsection{Minimization equations: just the Hartree - Fock approximation}

In fermionic systems one obviously does not have nonzero expectation values
for (on - shell, $J=0$) odd cumulants, namely $\left \langle X\right \rangle
=\left \langle X_{1}X_{2}X_{3}\right \rangle $ vanish on - shell. Unlike in
the bosonic theories this does not hinge on the preservation of symmetries.
As a consequence the first and the third minimization equations are
trivially satisfied. The gap equation on shell is (we do not mark ``tr" for
the variational on-shell green function in this section for the simplicity
of notation):

\begin{equation}
\Gamma ^{AB}=-\left[ G^{-1}\right] ^{AB}=-T^{AB}-\frac{1}{2}V^{ABXY}G^{XY}%
\text{.}  \label{gapeqferm}
\end{equation}%
The first (matrix) equality has a sign opposite to that for bosons. The
equation is just the Hartree - Fock (HF) self - consistency condition\cite%
{NO}. This means that the complexity of the only nonlinear operation within
the CCA scheme for fermions coincides with the complexity of a presumably
less precise CGA (equal for calculation of the one body correlator to the HF
approximation). The additional complexity arises only to the fact that
within CCA the connected correlation does not coincides with the truncated
correlator, as we have seen in the previous section and will be assessed
later.

Therefore we turn to the derivation of the correction $\Delta \Gamma ^{AB}$
formally similar to that in the bosonic model, Eq.(\ref{deltaGamdef}).

\subsubsection{Correction to correlator}

The CCA inverse correlator is derivative of the off - shell ES, Eq.(\ref%
{ESfermion}):%
\begin{eqnarray}
\Gamma _{\left( III\right) }^{AB} &=&\frac{\delta J^{A}}{\delta \psi ^{B}}%
=-T^{AB}-\frac{1}{2}\left( V^{ABX_{1}X_{2}}\psi ^{X_{1}}\psi
^{X_{2}}+V^{ABX_{1}X_{2}}G^{X_{1}X_{2}}-V^{AX_{2}X_{3}X_{4}}\psi
^{X_{2}}\Gamma ^{X_{1}B}\left \langle X_{1}X_{3}X_{4}\right \rangle \right)
\label{fullGamfer} \\
&&-\frac{1}{3!}V^{AX_{1}X_{2}X_{3}}\frac{\delta }{\delta \psi ^{B}}\left
\langle X_{1}X_{2}X_{3}\right \rangle \text{.}  \notag
\end{eqnarray}%
The on - shell nonzero contributions to $\Gamma _{\left( III\right) }^{AB}$
in the fermionic model are written via correction $\Gamma _{\left(
III\right) }^{AB}=\Gamma ^{AB}+\Delta \Gamma ^{AB}$ as:%
\begin{equation}
\Delta \Gamma ^{AB}=-\frac{1}{3!}V^{AX_{1}X_{2}X_{3}}\frac{\delta }{\delta
\psi ^{B}}\left \langle X_{1}X_{2}X_{3}\right \rangle =-\frac{1}{6}%
V^{AX_{1}X_{2}X_{3}}\left[ B|X_{1}X_{2}X_{3}\right] \text{.}  \label{delgam2}
\end{equation}%
The first term are just the truncated inverse correlator (or the covariant
gaussian inverse correlation) in view of the gap equation, Eq.(\ref%
{gapeqferm}).The chain, a derivative of truncated three - point connected
correlator will be denoted by
\begin{equation}
\frac{\delta }{\delta \psi ^{B}}\left \langle X_{1}X_{2}X_{3}\right \rangle
\equiv \left[ B|X_{1}X_{2}X_{3}\right] \text{.}  \label{fermchaindef}
\end{equation}%
The ``chain" is found from the derivative of the third DS equation, Eq.(\ref%
{thirdferm}).

\subsubsection{The chain equation}

Differentiating the ``connected version" of the third truncated DS equation,%
\begin{eqnarray}
\left \langle Z_{1}Z_{2}Z_{3}\right \rangle
&=&G^{Z_{1}Y_{1}}G^{Z_{2}Y_{2}}G^{Z_{3}Y_{3}}V^{Y_{1}Y_{2}Y_{3}X}\psi ^{X}+%
\frac{1}{2}V^{Y_{1}Y_{2}Y_{3}Y_{4}}G^{Z_{2}Y_{2}}G^{Z_{3}Y_{3}}\left \langle
Z_{1}Y_{1}Y_{4}\right \rangle  \label{thirdcon} \\
&&+\frac{1}{2}V^{Y_{1}Y_{2}Y_{3}Y_{4}}G^{Z_{1}Y_{1}}G^{Z_{2}Y_{2}}\left%
\langle Z_{3}Y_{3}Y_{4}\right \rangle \text{,}
\end{eqnarray}%
one obtains on - shell:%
\begin{eqnarray}
\left[ B|Z_{1}Z_{2}Z_{3}\right]
&=&G^{Z_{1}Y_{1}}G^{Z_{2}Y_{2}}G^{Z_{3}Y_{3}}V^{Y_{1}Y_{2}Y_{3}B}+\frac{1}{2}%
V^{Y_{1}Y_{2}Y_{3}Y_{4}}G^{Z_{2}Y_{2}}G^{Z_{3}Y_{3}}\left[ B|Z_{1}Y_{1}Y_{4}%
\right]  \label{chaineqferm} \\
&&+\frac{1}{2}V^{Y_{1}Y_{2}Y_{3}Y_{4}}G^{Z_{1}Y_{1}}G^{Z_{2}Y_{2}}\left[
B|Z_{3}Y_{3}Y_{4}\right] \text{.}  \notag
\end{eqnarray}%
One can prove that the chain is antisymmetric under $\left[ B|Z_{1}Z_{2}Z_{3}%
\right] =-\left[ B|Z_{3}Z_{2}Z_{1}\right] $ only. For example $\left[
B|Z_{1}Z_{2}Z_{3}\right] \neq -\left[ B|Z_{2}Z_{1}Z_{3}\right] $. This is
typical for ``truncated" (non - covariant) quantities that was observed
already in CGA\cite{GN,Wang17}$.$

The first important observation is that the chain equation is linear, as in
the bosonic case. An additional important observation is that the parameter $%
B$ is a ``spectator", so, since it is an external index in the correlator
itself, Eq.(\ref{delgam2}), one do not have to run over all its values.

\subsubsection{Most economic linear combination of chains: V-chains}

The chain equations although linear are very numerous. On the other hand, a
glance at the expression for the correction to the inverse correlator, Eq.(%
\ref{delgam2}),\ shows that only $N$ linear combinations are required. A
general question arises whether some linear combinations are ``closed" on
themselves. We have already noticed ``spectators" in Eq.(\ref{chaineqferm}).

After some ``trial and error", it turns out that the following combinations
of the chains are more convenient. Defining the convenient chains
combination as
\begin{equation}
\left \langle B|AY|X_{1}\right \rangle =V^{AYX_{2}X_{3}}\left[
B|X_{1}X_{2}X_{3}\right] \text{,}  \label{Vchain}
\end{equation}%
the correction becomes a ``trace":
\begin{equation}
\Delta \Gamma ^{AB}=-\frac{1}{6}V^{AX_{1}X_{2}X_{3}}\left[ B|X_{1}X_{2}X_{3}%
\right] =-\frac{1}{6}\left \langle B|AY|Y\right \rangle \text{,}
\label{delgamVch}
\end{equation}%
where $Y$ is the summation index, and $Y$ and $Z$ indices are also summation
indices in the equation below. The ``V - chain" (antisymmetric in the second
and third index) obeys the corresponding linear combination of the chain
equations:%
\begin{equation}
\left \langle B|X_{1}X_{2}|R\right \rangle =V^{X_{1}X_{2}Z_{1}Z_{2}}\left \{
\begin{array}{l}
\left \langle Z_{1}Y_{1}\right \rangle \left \langle RY_{3}\right \rangle
\left \langle Z_{2}Y_{2}\right \rangle V^{Y_{1}Y_{2}Y_{3}B}+\frac{1}{2}\left
\langle Z_{1}Y_{1}\right \rangle \\
\times \left( \left \langle Z_{2}Y_{2}\right \rangle \left \langle
B|Y_{1}Y_{2}|R\right \rangle -\left \langle RY_{2}\right \rangle \left
\langle B|Y_{1}Y_{2}|Z_{2}\right \rangle \right)%
\end{array}%
\right \} \text{.}  \label{Vchaineq}
\end{equation}%
This allows to solve the set of linear equations less times and in addition
to use the ``reduce" routines.

Let us now apply the rather abstract formalism to a sufficiently general
charge conserving electronic system.

\subsection{Charge conserving electron system with pairwise interaction}

\subsubsection{Down folding of the microscopic Hamiltonian. Measurement of
the electron correlator by photo - emission.}

Electronic system in a crystal on the microscopic level is described by a
many - body Hamiltonian with Coulomb interactions between electrons and very
slowly moving nuclei (considered typically as an external potential). The
problem of the determination of the electromagnetic, thermal and other
condensed matter properties that incorporate the excitation effects, is
divided into two steps, see Fig. 5. First DFT is used to calculate
coefficients of the effective low energy (meaning close to the Fermi level
within 1eV or so) \textquotedblleft down - folded\textquotedblright \ model%
\cite{Casula}. The bands far from the Fermi level and irrelevant parts of
the original Hilbert space are thus projected out.

Although there is a certain ambiguity what electronic model contains most of
the relevant information, there exists a reasonable choice for insulators
and semiconductors that still allows GW, DMFT, and as will be shown next,
CCA. The crucial novel feature of the CCA is its manifest charge
conservation (covariance).

The model is defined on a lattice with sufficiently large unit cell. The
downfolded spectrum is given by the two body electronic Greens function $%
G_{0}^{AB}(¦Ø ,p)$ with relevant bands indexed by $A=1,...,N_{b}$. The
number of retained atomic orbitals (including spin) $N_{b}$ should not be
not be too large for CCA (see estimates in the discussion section). The
typically rather long range unscreen Coulomb interactions are described
economically by a \textquotedblleft photon Greens function" $W_{0}^{AB}(%
¦Ø ,p)$. The restriction of the effective downfolded four fermion
interaction to being pairwise of the density - density type is not obvious.
CCA in general four - Fermi interaction is not feasible at presently
available computational power. To be successful the downfolding generally
utilizes the maximally localized DFT wavefunctions\cite{Vanderbilt}.

Recently the electronic Greens function is being ``measured" rather directly
by various photo - emission probes like ARPES - angle resolved photo -
emission spectroscopy. This is especially important for novel clean
crystalline materials, many of them low dimensional semiconductors and so
called Weyl semi - metals like $BN$, graphene etc. The Matsubara frequency
correlator calculated within CCA should therefore be analytically continued
to the spectral weight and density of states to be compared with experiments
and other methods, see Fig. 5. With this in mind a sufficiently general model%
\cite{NO,Guiliani} is defined formally next.
\begin{figure}[tbp]
\begin{center}
\includegraphics[width=16cm]{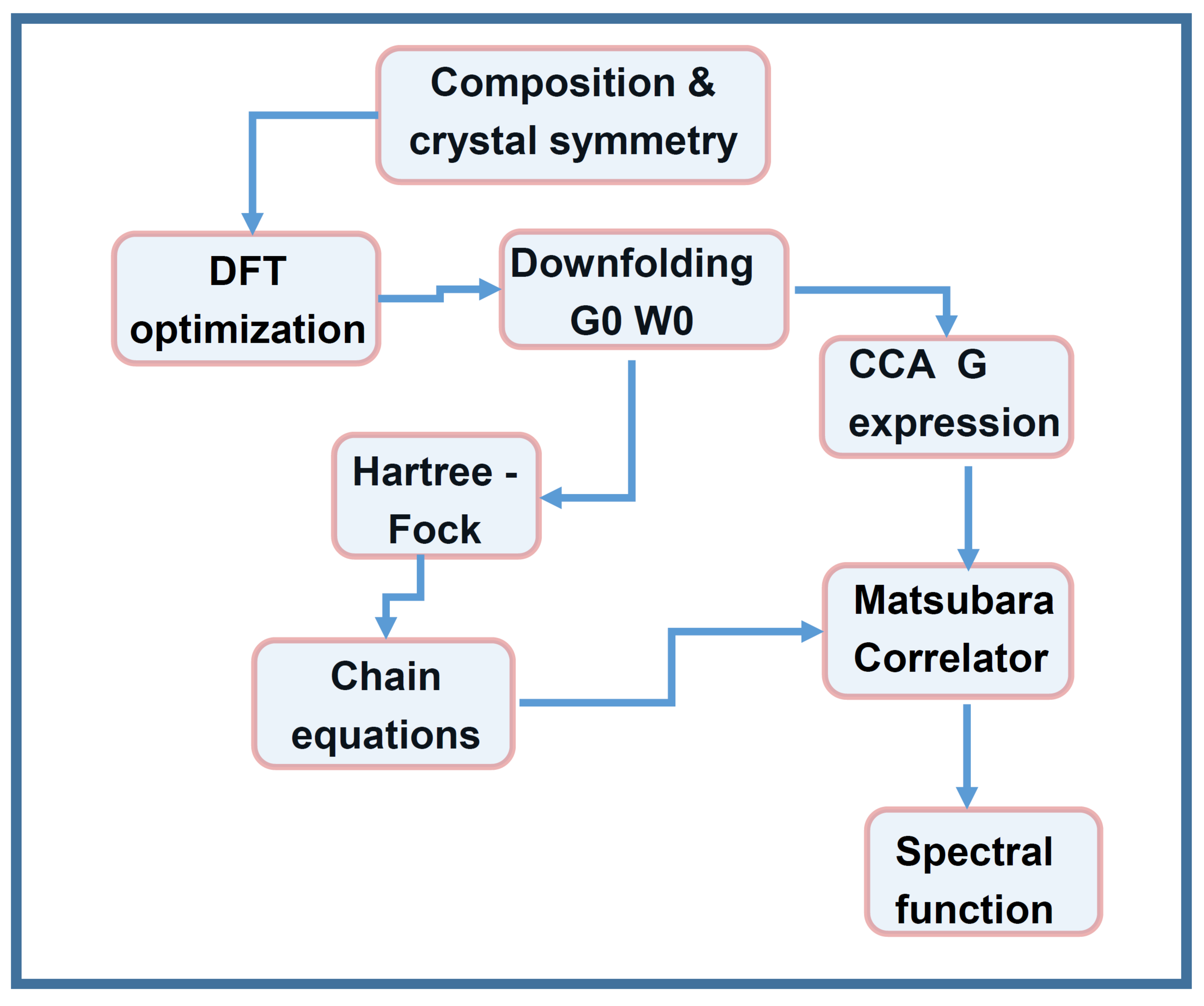}
\end{center}
\par
\vspace{-0.5cm}
\caption{Flow chart of the CCA calculation of the electron field correlator
of crystalline material.}
\end{figure}

\subsubsection{Matsubara action for the general downfolded model.}

The Matsubara action of the general pairwise interacting downfolded electron
model has a form\cite{NO}:%
\begin{eqnarray}
\mathcal{A}\left[ \psi \right] &=&\psi _{a}^{\ast }T_{ab}\psi _{b}^{\cdot }+%
\frac{1}{2}\psi _{a}^{\ast }\psi _{a}^{\cdot }V_{ab}\psi _{b}^{\ast }\psi
_{b}^{\cdot };  \label{chargeconserving} \\
Z\left[ J\right] &=&\int D\psi D\psi ^{\ast }\exp \left[ -\mathcal{A}\left[
\psi \right] +J_{a}^{\ast }\psi _{a}^{\ast }+J_{a}^{\cdot }\psi _{a}^{\cdot }%
\right] \text{.}  \notag
\end{eqnarray}%
\  \  \  \  \  \  \  \  \  \  \  \  \  \  \  \  \  \  \  \  \  \  \  \  \  \  \  \  \  \  \  \  \  \  \  \  \  \
\  \  \  \  \  \  \  \  \  \  \  \  \  \  \  \  \  \  \  \  \  \  \  \  \  \  \  \  \  \  \  \  \  \  \  \  \  \
\  \  \  \  \  \  \  \  \  \  \  \  \  \  \  \  \  \  \  \  \  \  \  \  \ Charge conservation is
explicit here. This should be matched with fully symmetrized Grassmanian
form, Eq.(\ref{fermionaction}):

\begin{equation}
\mathcal{A}=\frac{1}{2}\psi _{a}^{A}T_{ab}^{AB}\psi _{B}+\frac{1}{4!}%
V_{abcd}^{ABCD}\psi _{a}^{A}\psi _{b}^{B}\psi _{c}^{C}\psi
_{d}^{D}-J_{a}^{A}\psi _{a}^{A}\text{.}  \label{Nambuform}
\end{equation}%
Here $A=\ast ,\cdot $ is the charge (Nambu) index that originate from the
creation and annihilation operators in the many - body Hamiltonian\cite{NO}.
The rest of the indices are contained in $a=\left \{
position,time,band\right \} $. The ``band" index includes spin. The
interaction is of the density - density form and thus $V_{ab}=V_{ba}$. The
result is:%
\begin{eqnarray}
T_{ab}^{AB} &=&\delta ^{A\ast }\delta ^{B\cdot }T_{ab}-\delta ^{A\cdot
}\delta ^{B\ast }T_{ba};  \label{TV} \\
V_{y_{1}y_{2}y_{3}y_{4}}^{Y_{1}Y_{2}Y_{3}Y_{4}} &=&\left \{
\begin{array}{c}
\delta _{y_{3}y_{4}}V_{y_{3}y_{1}}\delta _{y_{1}y_{2}}\left( \delta ^{\ast
Y_{4}}\delta ^{\cdot Y_{3}}-\delta ^{\cdot Y_{4}}\delta ^{\ast Y_{3}}\right)
\left( \delta ^{\ast Y_{2}}\delta ^{\cdot Y_{1}}-\delta ^{\cdot Y_{2}}\delta
^{\ast Y_{1}}\right) \\
-\delta _{y_{2}y_{4}}V_{y_{2}y_{1}}\delta _{y_{1}y_{3}}\left( \delta ^{\ast
Y_{4}}\delta ^{\cdot Y_{2}}-\delta ^{\cdot Y_{4}}\delta ^{\ast Y_{2}}\right)
\left( \delta ^{\ast Y_{3}}\delta ^{\cdot Y_{1}}-\delta ^{\cdot Y_{3}}\delta
^{\ast Y_{1}}\right) \\
-\delta _{y_{1}y_{4}}V_{y_{3}y_{1}}\delta _{y_{3}y_{2}}\left( \delta ^{\ast
Y_{4}}\delta ^{\cdot Y_{1}}-\delta ^{\cdot Y_{4}}\delta ^{\ast Y_{1}}\right)
\left( \delta ^{\ast Y_{2}}\delta ^{\cdot Y_{3}}-\delta ^{\cdot Y_{2}}\delta
^{\ast Y_{3}}\right)%
\end{array}%
\right \} \text{.}  \notag
\end{eqnarray}%
This model is now amenable to the CCA approximation scheme. In the present
paper only the correlator (one body correlator or green function) is
computed.

\subsubsection{CCA equations for the correlator}

For charge unbroken case (no superconductivity) some simplifications occur.
All the ``charges" correlators like $G_{ab}^{\ast \ast }$ vanish on - shell.
The only correlators remaining are $G_{ab}^{\ast \cdot }\equiv G_{ab},\Gamma
_{ab}^{\ast \cdot }\equiv \Gamma _{ab}$, related by $\Gamma
_{ax}G_{bx}=-\delta _{ab}$ where $x$ is the summation index. General Nambu
gap equation, Eq.(\ref{gapeqferm}) in an electric charge preserving system
takes a form:%
\begin{equation}
\Gamma _{ab}=-T_{ab}-\delta _{ab}V_{xa}G_{xx}+V_{ba}G_{ba}\text{.}
\label{HFeq}
\end{equation}%
The chain correction to the inverse propagator, Eq.(\ref{deltaGamdef}) in
our case becomes,%
\begin{equation}
\Delta \Gamma _{ab}=\frac{1}{6}\left( \left \langle _{b}^{\cdot
}|_{xa}^{\cdot \ast }|_{x}^{\cdot }\right \rangle +\left \langle _{b}^{\cdot
}|_{xa}^{\ast \ast }|_{x}^{\ast }\right \rangle \right) \text{.}
\label{delGamreal}
\end{equation}

Only two charge components of the chain appear here due to the antisymmetry
of the chain. Let us denote them as diffuson and cooperon chains in analogy
to similar expressions in diagrammatic many - body physics\cite{Akkermans}:

\begin{equation}
D_{bx_{1}x_{2}r}\equiv \left \langle _{b}^{\cdot }|_{x_{1}x_{2}}^{\cdot \ast
}|_{r}^{\cdot }\right \rangle ;\text{ \  \ }C_{bx_{1}x_{2}r}\equiv \left
\langle _{b}^{\cdot }|_{x_{1}x_{2}}^{\ast \ast }|_{r}^{\ast }\right \rangle
\text{.\ }  \label{cooperondef}
\end{equation}%
For these two quantities the general chain equation, Eq.(\ref{chaineqferm}),
closes:%
\begin{eqnarray}
D_{zx_{1}x_{2}r} &=&2V_{y_{1}z}\left \langle x_{1}y_{1}\right \rangle \left(
\left \langle y_{1}r\right \rangle \left \langle zx_{2}\right \rangle -\left
\langle zr\right \rangle \left \langle y_{1}x_{2}\right \rangle \right)
V_{x_{2}x_{1}}-2\delta _{x_{1}x_{2}}V_{y_{1}z}V_{x_{2}y_{2}}\left \langle
y_{2}y_{1}\right \rangle \left( \left \langle y_{1}r\right \rangle \left
\langle zy_{2}\right \rangle -\left \langle zr\right \rangle \left \langle
y_{1}y_{2}\right \rangle \right)  \label{Dchain} \\
&&+V_{x_{2}x_{1}}\left( -\left \langle x_{1}y_{1}\right \rangle \left
\langle y_{2}x_{2}\right \rangle D_{zy_{1}y_{2}r}+\frac{1}{2}\left \langle
x_{1}y_{1}\right \rangle \left \langle y_{2}r\right \rangle
D_{zy_{1}y_{2}x_{2}}+\frac{1}{2}\left \langle y_{1}x_{2}\right \rangle \left
\langle y_{2}r\right \rangle C_{zy_{1}y_{2}x_{1}}\right)  \notag \\
&&+\delta _{x_{1}x_{2}}V_{x_{1}y_{3}}\left( \left \langle y_{3}y_{1}\right
\rangle \left \langle y_{2}y_{3}\right \rangle D_{zy_{1}y_{2}r}-\frac{1}{2}%
\left \langle y_{3}y_{1}\right \rangle \left \langle y_{2}r\right \rangle
D_{zy_{1}y_{2}y_{3}}-\frac{1}{2}\left \langle y_{1}y_{3}\right \rangle \left
\langle y_{2}r\right \rangle C_{zy_{1}y_{2}y_{3}}\right) \text{;}  \notag \\
C_{bx_{1}x_{2}r} &=&2V_{x_{2}x_{1}}V_{by}\left \langle ry\right \rangle
\left( \left \langle bx_{2}\right \rangle \left \langle yx_{1}\right \rangle
-\left \langle bx_{1}\right \rangle \left \langle yx_{2}\right \rangle
\right) +V_{x_{2}x_{1}}\left( \frac{1}{2}\left \langle y_{2}x_{1}\right
\rangle \left \langle ry_{1}\right \rangle D_{by_{1}y_{2}x_{2}}\right.
\notag \\
&&\left. -\frac{1}{2}\left \langle y_{2}x_{2}\right \rangle \left \langle
ry_{1}\right \rangle D_{by_{1}y_{2}x_{1}}-\left \langle y_{2}x_{2}\right
\rangle \left \langle y_{1}x_{1}\right \rangle C_{by_{1}y_{2}r}\right) \text{%
.}  \notag
\end{eqnarray}%
Solution and application of these equations greatly simplifies when the
translational symmetry is utilized.

\subsubsection{Translation invariance}

In addition to charge conservation we assume the electronic system to be
invariant under a crystalline translation symmetry and time translations.
The model of relevant number of $N_{f}$ bands (including spin) constructed
on the lattice with periodic boundary conditions $N_{s}$ in each direction,
to keep notations as simple as possible, the square lattice is assumed with
lattice spacing defining the unit of length $a=1$. The points therefore are $%
r_{i}=1,..N_{s}$, $i=1,..,D$ (dimensionality). At temperature $T$ the
Matsubara (Euclidean) time is also discretized $t=1,...N_{t}$ in the range $%
0<\frac{1}{TN_{t}}t\leq 1/T$ and $\psi _{t}$ is antiperiodic\cite{NO}.

Therefore the electron field is carrying two types of indices $a=\left \{ A,%
\mathbf{a}\right \} $, the band index will be consistently written as a
superscript, while the space - time index $\mathbf{a}$ will be eventually
substituted by integer valued wave number $k$ and the Matsubara frequency $n$%
, so that we map $\psi _{a}^{\ast }\rightarrow \psi _{a}^{A\ast }$.
Definitions of the discrete Fourier transform (FT) of the complex
grassmanian field is%
\begin{equation}
\psi _{\mathbf{a}}^{A\ast }=\sqrt{\frac{T}{N_{s}^{D}}}\sum%
\nolimits_{k_{1},...k_{D}=1}^{N_{s}}\sum \nolimits_{n=1}^{M}exp\left[ -2\pi
i\left( \frac{\left( n+1/2\right) t}{N_{t}}+\frac{k_{i}r_{i}}{N_{s}}\right) %
\right] \psi _{\alpha }^{A\ast }\text{.}  \label{FTdef}
\end{equation}

Now $\alpha =\left \{ n,k_{1},...,k_{D}\right \} $ enumerates the space -
time components of the energy-momentum basis. Translation invariance (energy
and momentum conservation) leads to the following FT for the correlators:
\begin{equation}
G_{ab}^{AB}=\frac{T}{N_{s}^{D}}\sum \nolimits_{\alpha }\exp \left[ i\left(
\mathbf{b-a}\right) \cdot \mathbf{\alpha }\right] g_{\alpha }^{AB}\text{,}
\label{Gft}
\end{equation}%
where $\mathbf{\alpha }=2\pi \left \{ \frac{\left( n+1/2\right) }{N_{t}},%
\frac{k}{N_{s}}\right \} $. For the inverse propagator it is convenient to
define FT by

\begin{equation}
\Gamma _{ab}^{AB}=\frac{\tau }{N_{t}N_{s}^{D}}\sum \nolimits_{\alpha }\exp %
\left[ i\left( \mathbf{a-b}\right) \cdot \mathbf{\alpha }\right] \gamma
_{\alpha }^{BA}\text{,}  \label{gammaft}
\end{equation}%
where $\tau =\frac{1}{TN_{t}}$ is the Matsubara time step, so that $g_{%
\mathbf{\alpha }}^{BX}\gamma _{\mathbf{\alpha }}^{XA}=\delta ^{AB}$.
Consequently tunneling and interaction potentials FT are,%
\begin{eqnarray}
T_{ab}^{AB} &=&\frac{\tau }{N_{t}N_{s}^{D}}\sum \nolimits_{\alpha }\exp %
\left[ i\left( \mathbf{a-b}\right) \cdot \mathbf{\alpha }\right] t_{\alpha
}^{BA}\text{;}  \label{FT} \\
V_{ab}^{AB} &=&\frac{\tau }{N_{t}N_{s}^{D}}\sum \nolimits_{\mathbf{\lambda }%
}\exp \left[ i\left( \mathbf{a-b}\right) \cdot \mathbf{\lambda }\right]
v_{\lambda }^{AB}\text{,}  \notag
\end{eqnarray}%
where $\mathbf{\lambda }=2\pi \left \{ \frac{n}{N_{t}},\frac{l}{N_{s}}%
\right
\} $ has bosonic Matsubara frequency.

Using these definitions the HF equation, Eq.(\ref{HFeq}), becomes:

\begin{equation}
\gamma _{\alpha }^{BA}=-t_{\alpha }^{BA}-\frac{T}{N_{s}^{D}}\sum
\nolimits_{\chi }\left( \delta ^{AB}v_{0}^{XA}g_{\chi }^{XX}-v_{\alpha -\chi
}^{AB}g_{\chi }^{BA}\right) \text{.}  \label{fouriergapeq}
\end{equation}%
The correction to the inverse propagator takes a form

\begin{equation}
\Delta \gamma _{\alpha }^{BA}=\frac{1}{6}\frac{T}{N_{s}^{D}}\sum
\nolimits_{\kappa }\left( d_{\alpha \kappa \alpha }^{BXAX}+c_{\alpha \kappa
\alpha }^{BXAX}\right) \text{,}  \label{corrFT}
\end{equation}%
where$\ $
\begin{eqnarray}
D_{ax_{1}x_{2}r} &=&\left( N_{s}^{D}N_{t}\right) ^{-3}\sum \nolimits_{\alpha
\kappa \gamma }exp\left[ -\left( a\alpha -x_{1}\left( \kappa -\gamma \right)
-x_{2}\gamma +\left( \kappa -\zeta \right) r\right) \right] d_{\alpha \kappa
\gamma }\text{;}  \label{chainft} \\
C_{ax_{1}x_{2}r} &=&\left( N_{s}^{D}N_{t}\right) ^{-3}\sum \nolimits_{\alpha
\kappa \gamma }ixp\left[ -\left( a\alpha -x_{1}\left( \kappa -\gamma \right)
-x_{2}\gamma +\left( \kappa -\zeta \right) r\right) \right] c_{\alpha \kappa
\gamma }\text{.}  \notag
\end{eqnarray}

Finally the chain equation are (the spectator frequency-wave-vector indices
are $\alpha $ and $B$)%
\begin{eqnarray}
c_{\alpha ,\kappa ,\gamma }^{BX_{1}X_{2}R} &=&\frac{2T}{N_{s}^{D}}\left(
v_{\alpha +\chi -\gamma }^{X_{1}X_{2}}v_{\chi }^{BY}g_{\kappa -\alpha
}^{RY}g_{\alpha +\chi }^{BX_{2}}g_{\kappa -\alpha -\chi }^{YX_{1}}-v_{\kappa
-\alpha -\chi -\gamma }^{X_{1}X_{2}}v_{\chi }^{BY}g_{\kappa -\alpha
}^{RY}g_{\alpha +\chi }^{BX_{1}}g_{\kappa -\alpha -\chi }^{YX_{2}}\right) +%
\frac{T}{2N_{s}^{D}}\times  \label{chainFT} \\
&&\left( v_{\alpha -\gamma -\chi }^{X_{1}X_{2}}g_{\chi +\kappa -\alpha
}^{Y_{2}X_{1}}g_{\kappa -\alpha }^{RY_{1}}d_{\alpha ,\chi ,\chi +\kappa
-\alpha }^{BY_{1}Y_{2}X_{2}}-v_{\kappa +\chi -\alpha -\gamma
}^{X_{1}X_{2}}g_{\kappa +\chi -\alpha }^{Y_{2}X_{2}}g_{\kappa -\alpha
}^{RY_{1}}d_{\alpha ,\chi ,\kappa +\chi -\alpha
}^{BY_{1}Y_{2}X_{1}}-2v_{\chi -\kappa _{2}}^{X_{1}X_{2}}g_{\chi
}^{Y_{2}X_{2}}g_{\kappa -\chi }^{Y_{1}X_{1}}c_{\alpha ,\kappa ,\chi
}^{BY_{1}Y_{2}R}\right) ;  \notag \\
d_{\alpha ,\kappa ,\gamma }^{BX_{1}X_{2}R} &=&\frac{2T}{N_{s}^{D}}\left(
v_{\chi -\gamma }^{X_{1}X_{2}}v_{\alpha -\chi }^{Y_{1}Z}g_{\chi -\kappa
}^{X_{1}Y_{1}}g_{\alpha -\kappa }^{Y_{1}R}g_{\chi }^{BX_{2}}-v_{\chi -\gamma
}^{X_{1}X_{2}}v_{\kappa }^{Y_{1}B}g_{\chi -\kappa }^{X_{1}Y_{1}}g_{\alpha
-\kappa }^{BR}g_{\chi }^{Y_{1}X_{2}}\right)  \notag \\
&&+\frac{2T}{N_{s}^{D}}\delta ^{X_{1}X_{2}}v_{\kappa }^{Y_{1}B}g_{\chi
-\kappa }^{Y_{2}Y_{1}}\left( v_{\kappa }^{X_{2}Y_{2}}g_{\alpha -\kappa
}^{BR}g_{\chi }^{Y_{1}Y_{2}}-v_{\alpha -\chi }^{X_{2}Y_{2}}g_{\alpha -\kappa
}^{Y_{1}R}g_{\chi }^{BY_{2}}\right) +\frac{T}{2N_{s}^{D}}  \notag \\
&&\times \left( -2v_{\chi -\gamma }^{X_{1}X_{2}}g_{\chi -\kappa
}^{X_{1}Y_{1}}g_{\chi }^{_{Y_{2}X_{2}}}d_{\alpha ,\kappa ,\chi
}^{BY_{1}Y_{2}R}+v_{\alpha -\chi -\gamma }^{X_{1}X_{2}}g_{\alpha -\chi
-\kappa }^{X_{1}Y_{1}}g_{\alpha -\kappa }^{Y_{2}R}d_{\alpha ,\chi ,\alpha
-\kappa }^{BY_{1}Y_{2}X_{2}}+v_{\chi -\gamma }^{X_{1}X_{2}}g_{\chi
}^{Y_{1}X_{2}}g_{\alpha -\kappa }^{Y_{2}R}c_{\alpha ,\alpha -\kappa ,\alpha
-\kappa }^{BY_{1}Y_{2}X_{1}}\right)  \notag \\
&&+\frac{T}{2N_{s}^{D}}\delta ^{X_{1}X_{2}}v_{\kappa }^{X_{1}Y_{3}}\left(
2g_{\chi -\kappa }^{Y_{3}Y_{1}}g_{\chi }^{Y_{2}Y_{3}}d_{\alpha ,\kappa ,\chi
}^{BY_{1}Y_{2}R}-g_{\alpha -\kappa -\chi }^{Y_{3}Y_{1}}g_{\alpha -\kappa
}^{Y_{2}R}d_{\alpha ,\chi ,\alpha -\kappa }^{BY_{1}Y_{2}Y_{3}}-g_{\kappa
+\chi -\alpha }^{Y_{1}Y_{3}}g_{\alpha -\kappa }^{Y_{2}R}c_{\alpha ,\chi
,\alpha -\kappa }^{BY_{1}Y_{2}Y_{3}}\right) \text{,}  \notag
\end{eqnarray}%
with $\chi $ being the only summation index. To test the CCA in a fermionic
model, one should apply the method to an exactly solvable one. Exact
solution exists for sufficiently small $N_{s}$ in the case of local
interaction - Hubbard model. We therefore apply CCA to the case of Hubbard
model and compare it to the exact diagonalization\cite{ED} (ED) in the case
of $N_{s}=1$ (quantum dot) and 1D with small finite $N_{s}$.

$\  \ $

\section{Fermionic benchmarks: quantum dot and one dimensional Hubbard model.%
}

\subsection{The CCA approximation in D - dimensional one band Hubbard model.}

\subsubsection{The model, gap and chain equations.}

The single band Hubbard model is defined on the $D$ dimensional hypercubic
lattice. The tunneling amplitude to the neighbouring site in any direction $%
i=1,...,D$ is denoted in literature by $t$. We chose it to be the unit of
energy $t=1$. Similarly the lattice spacing sets the unit of length $a=1$
and $\hbar =1$. The Hamiltonian is:

\begin{equation}
H=\sum \nolimits_{r_{1,...},r_{D}=1}^{N_{s}}\left \{ -\sum
\nolimits_{i}\left( a_{r}^{A\dagger }a_{r+\widehat{i}}^{A}+h.c.\right) -\mu
n_{r}-ha_{r}^{A\dagger }\sigma _{z}^{AB}a_{r}^{B}+Un_{r}^{\upharpoonleft
}n_{r}^{\downarrow }\right \} \text{.}  \label{HubbardH}
\end{equation}%
The chemical potential $\mu $ and the on - site repulsion energy $U$ are
therefore given in units of the hopping energy The ``band" index therefore
takes two values $A,B=\uparrow ,\downarrow $. The hopping direction is
denoted by $\widehat{i}$ as in statistical physics model\cite{Rothe,Kleinert}
of Eq.(\ref{latLaplacian}). The density and its spin components are $%
n_{r}=n_{r}^{\upharpoonleft }+n_{r}^{\downarrow }$ with $n_{r}^{A}\equiv
a_{r}^{A\dagger }a_{r}^{A}$. External magnetic field $h$ makes the electrons
polarized. At half filling $\mu =\frac{U}{2}$.

The discretized Matsubara action is\cite{NO},
\begin{equation}
\mathcal{A}=\tau \sum \nolimits_{t,r}\left \{
\begin{array}{c}
\frac{1}{\tau }\left( \psi _{t+1,r}^{A\ast }\psi _{t,r}^{A}-\psi
_{t,r}^{A\ast }\psi _{t,r}^{A\ast }\right) \ -\frac{1}{2}\sum
\nolimits_{i}\left( \psi _{t,r}^{A\dagger }\psi _{t,r+\widehat{i}}^{A}+\psi
_{t,x}^{A\dagger }\psi _{t,r-\widehat{i}}^{A}\right) \\
-\left( \mu _{H}-\frac{U}{2}\right) n_{r}-h\psi _{r}^{A\dagger }\sigma
_{z}^{AB}\psi _{r}^{B}-U\psi _{t,r}^{\upharpoonleft \ast }\psi
_{t,r}^{\downarrow \ast }\psi _{t,r}^{\upharpoonleft }\psi
_{t,r}^{\downarrow }%
\end{array}%
\right \} \text{,}  \label{Liuaction}
\end{equation}%
where $n_{t,r}\equiv \psi _{t,r}^{X\dagger }\psi _{t,r}^{X}$ and the ``slice
size" in anti - periodic Matsubara time is $\tau =\left( TN_{t}\right) ^{-1}$%
, where $T$ is temperature and $N_{t}$ is the number of points in the
compact time axis\cite{NO}. Therefore the hopping matrix in
frequency-momentum space of the corresponding Matsubara action, Eq.(\ref%
{chargeconserving}) is
\begin{eqnarray}
t_{n,k}^{AB} &=&\delta ^{AB}t_{n,k}-h\sigma _{z}^{AB};\text{ \ }
\label{hoppingtHub} \\
t_{n,k} &=&\frac{1}{\tau }\left( \exp \left[ i\frac{2\pi \left( n+1/2\right)
}{N_{t}}\right] -1\right) -2\sum \nolimits_{i}\cos \left[ \frac{2\pi k_{i}}{%
N_{s}}\right] -\mu _{H}\text{,}  \notag
\end{eqnarray}%
while the interaction is just a constant:%
\begin{equation}
v_{nk}^{AB}=U\text{.}  \label{Hubbar_int}
\end{equation}

The gap equation, Eq.(\ref{HFeq}), in this case takes a form:
\begin{eqnarray}
\gamma _{\zeta }^{BA} &=&-t_{\zeta }^{AB}-U\left( \delta
^{AB}n^{XX}-n^{BA}\right) \text{;}  \label{gapeqHub} \\
n^{AB} &=&\frac{T}{N_{s}^{D}}\sum \nolimits_{\chi }g_{\chi }^{AB}\text{,}
\notag
\end{eqnarray}%
where the $D+1$ dimensional notations like $\zeta =\left \{ n,k\right \} $
will be used to simplify the expressions. In the range of parameters
considered in the present paper the spin rotation $U\left( 1\right) $
symmetry along the z axis will be assumed unbroken (A larger $SU\left(
2\right) $ symmetry appears at zero magnetic field\cite{Korepin}, which is
not discussed here since we focus initially on the ``symmetry" unbroken
phases). The most Ansatz is $n^{\uparrow \uparrow }=n^{\uparrow };$ $%
n^{\downarrow \downarrow }=n^{\downarrow };$ $n^{\uparrow \downarrow
}=n^{\downarrow \uparrow }=0$.

Therefore the only two nontrivial diagonal components (denoted by $g_{\zeta
}^{AA}\equiv g_{\zeta }^{A}$) of the truncated inverse propagator are

\begin{equation}
1/g_{\zeta }^{A}=-t_{\zeta }^{A}-Un^{\overline{A}}\text{,}  \label{HubbardHF}
\end{equation}%
where the bar over the spin index means that the spin was flipped. The
couple of algebraic self consistent equations finally is:

\begin{equation}
n^{A}=-\frac{T}{N_{s}^{D}}\sum \nolimits_{\chi }\frac{1}{t_{\chi }^{A}+Un^{%
\overline{A}}}\text{,}  \label{HubHF}
\end{equation}%
and is easily solved numerically.

The set of chain equations greatly simplified exactly as in the bosonic
model since the interaction is local. One requires only coincident
coordinates for the cooperon $C_{zxxr}^{ZX_{1}X_{2}R}$ and diffuson $%
D_{zxxr}^{ZX_{1}X_{2}R}$ defined in Eq.(\ref{cooperondef}). Moreover the
remaining $U\left( 1\right) $ spin symmetry (spin along the z axis) limits
the nonzero spin components. Generally Pauli symmetry demands that for
cooperon with fixed spectator spin $Z$, the only nonzero choice is $C=C^{ZZ%
\overline{Z}\overline{Z}}$. For $D$ the symmetry leaves three choices $%
D^{1}=D^{ZZZZ}$, $D^{2}=D^{Z\overline{Z}\overline{Z}Z}$ and $D^{3}=D^{Z%
\overline{Z}Z\overline{Z}}$. The resulting set of chain equations in Fourier
space is:

\begin{eqnarray}
\frac{N_{s}^{D}}{UT}c_{\zeta \kappa } &=&-2Ug_{\kappa -\zeta }^{\overline{Z}%
}g_{\kappa -\eta }^{Z}g_{\eta }^{\overline{Z}}+\frac{1}{2}g_{\kappa -\zeta
+\chi }^{Z}g_{\kappa -\zeta }^{\overline{Z}}d_{\zeta \chi }^{3}-\frac{1}{2}%
g_{\kappa -\zeta +\chi }^{\overline{Z}}g_{\kappa -\zeta }^{\overline{Z}%
}d_{\zeta \chi }^{2}-g_{\kappa -\eta }^{\overline{Z}}g_{\eta }^{Z}c_{\zeta
\kappa };  \label{chainHub} \\
\frac{N_{s}^{D}}{UT}d_{\zeta \kappa }^{1} &=&2Ug_{\zeta -\kappa }^{Z}g_{\eta
-\kappa }^{\overline{Z}}g_{\eta }^{\overline{Z}}+\frac{1}{2}g_{\kappa -\zeta
+\chi }^{\overline{Z}}g_{\zeta -\kappa }^{Z}c_{\zeta \chi }+g_{\eta -\kappa
}^{\overline{Z}}g_{\eta }^{\overline{Z}}d_{\zeta \kappa }^{2}-\frac{1}{2}%
g_{\zeta -\kappa -\chi }^{\overline{Z}}g_{\zeta -\kappa }^{Z}d_{\zeta \chi
}^{3};  \notag \\
\frac{N_{s}^{D}}{UT}d_{\zeta \kappa }^{2} &=&g_{\eta }^{Z}g_{\kappa +\eta
}^{Z}d_{\zeta \kappa }^{1}-\frac{1}{2}g_{\zeta -\kappa -\chi }^{Z}g_{\zeta
-\kappa }^{Z}d_{\zeta \chi }^{1};  \notag \\
\frac{N_{s}^{D}}{UT}d_{\zeta \kappa }^{3} &=&2Ug_{\eta -\kappa }^{\overline{Z%
}}g_{\zeta -\kappa }^{\overline{Z}}g_{\eta }^{Z}+\frac{1}{2}g_{-\zeta
+\kappa +\chi }^{Z}g_{\zeta -\kappa }^{\overline{Z}}c_{\zeta \chi }+\frac{1}{%
2}g_{\zeta -\kappa -\chi }^{\overline{Z}}g_{\zeta -\kappa }^{\overline{Z}%
}d_{\zeta \chi }^{2}-g_{\eta }^{\overline{Z}}g_{\eta +\kappa }^{Z}d_{\zeta
\kappa }^{3}\text{.}  \notag
\end{eqnarray}%
Here summation over bosonic ($\chi $) and fermionic ($\eta $)
frequencies/momenta are assumed.

The CCA correlator, Eq.(\ref{corrFT}) in this case is:
\begin{equation}
\Delta \gamma _{\zeta }^{ZZ}=\frac{T}{6N_{s}^{D}}\sum \nolimits_{\kappa
}\left( d_{\zeta \kappa }^{1}+d_{\zeta \kappa }^{3}-c_{\zeta \kappa }\right)
\text{.}  \label{corrHub}
\end{equation}%
It was calculated (using C++ program on parallel computer) for the cases of
the toy model (quantum dot) $D=0$ and $D=1$ for sufficiently small $N_{s}$
so that exact diagonalization is possible.

\begin{figure}[tbp]
\begin{center}
\includegraphics[width=18cm]{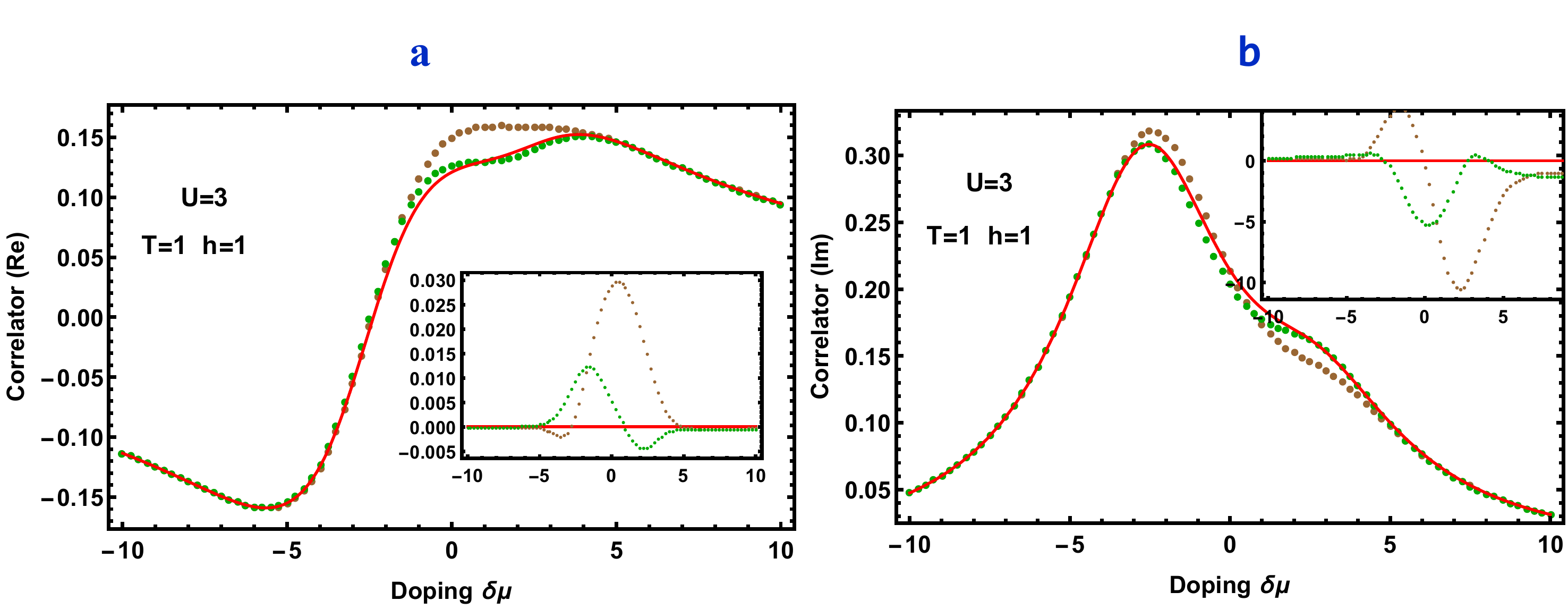}
\end{center}
\par
\vspace{-0.5cm}
\caption{Fig. 6. Real (a) and imaginary (b) parts of the Matsubara
correlator for quantum dot at intermediate value of the coupling.}
\end{figure}

\begin{figure}[tbp]
\begin{center}
\includegraphics[width=18cm]{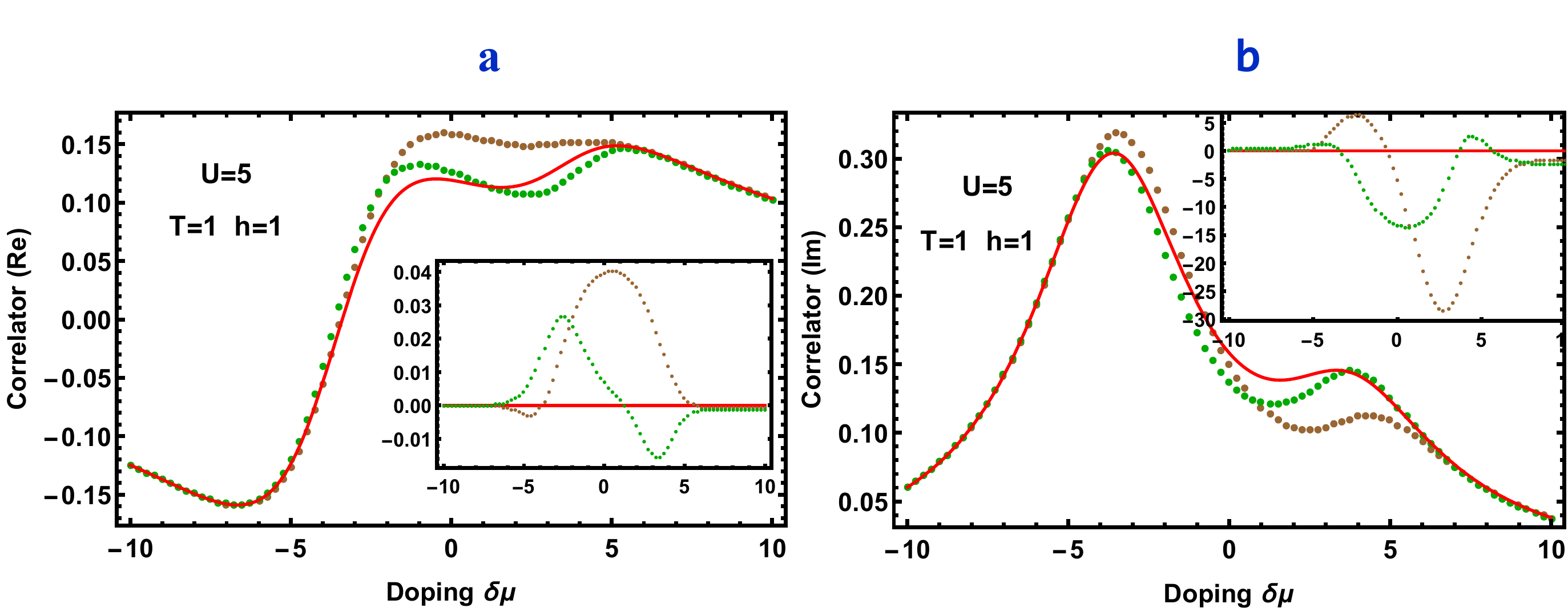}
\end{center}
\par
\vspace{-0.5cm}
\caption{Fig.7. Same as Fig. 6 for a rather strong coupling.}
\end{figure}

\subsection{Fermionic toy model: quantum dot}

Let us first consider an exactly solvable model of just a single site
(``quantum dot"). Recently artificial systems like that with several sited
Hubbard model were manifactured\cite{qdexp} and the experimental results
were compared with exact diagonalization (ED). The space indices are absent
in the $D=0$ model, so that the space-time index $\alpha $ stands for the
frequency after Fourier transform. The model can be solved with the result
for the correlator of the up spin being:

\begin{eqnarray}
g_{n}^{\uparrow } &=&-\frac{1}{Z}\left \{ \frac{1+e^{\left( \mu +h\right) /T}%
}{i\pi T\left( 2n+1\right) -\mu -h}+\frac{e^{\left( \mu -h\right)
/T}+e^{\left( 2\mu -U\right) /T}}{i\pi T\left( 2n+1\right) -\mu -h+U}\right
\} ;  \label{QDex} \\
Z &=&1+e^{\left( \mu +h\right) /T}+e^{\left( \mu -h\right) /T}+e^{\left(
2\mu -U\right) /T}.  \notag
\end{eqnarray}

This is presented as a red lines in Fig. 6. In Fig. 6a the real part of the
Matsubara correlator in wide range of doping $\delta \mu \equiv \mu -U/2$
for $U=3$ is given, while Fig. 6b exhibits the imaginary part. Temperature
and magnetic field were fixed at $T=1$, $h=1$. One observes a very good
agreement not only in perturbative domains for large absolute value of $%
\delta \mu $ (far away from half filling. The maximal deviations are $0.03$,
$0.01$ (real part) and $10\%$, $5\%$ (imaginary part) for CGA and CCA
respectively, see insets. In Fig. 7 the same is given for a strong coupling $%
U=5$. The agreement is worse, by still have maximal deviations are $0.04$, $%
0.025$ (real part) and $28\%$, $14\%$ for CGA and CCA.

\begin{figure}[tbp]
\begin{center}
\includegraphics[width=18cm]{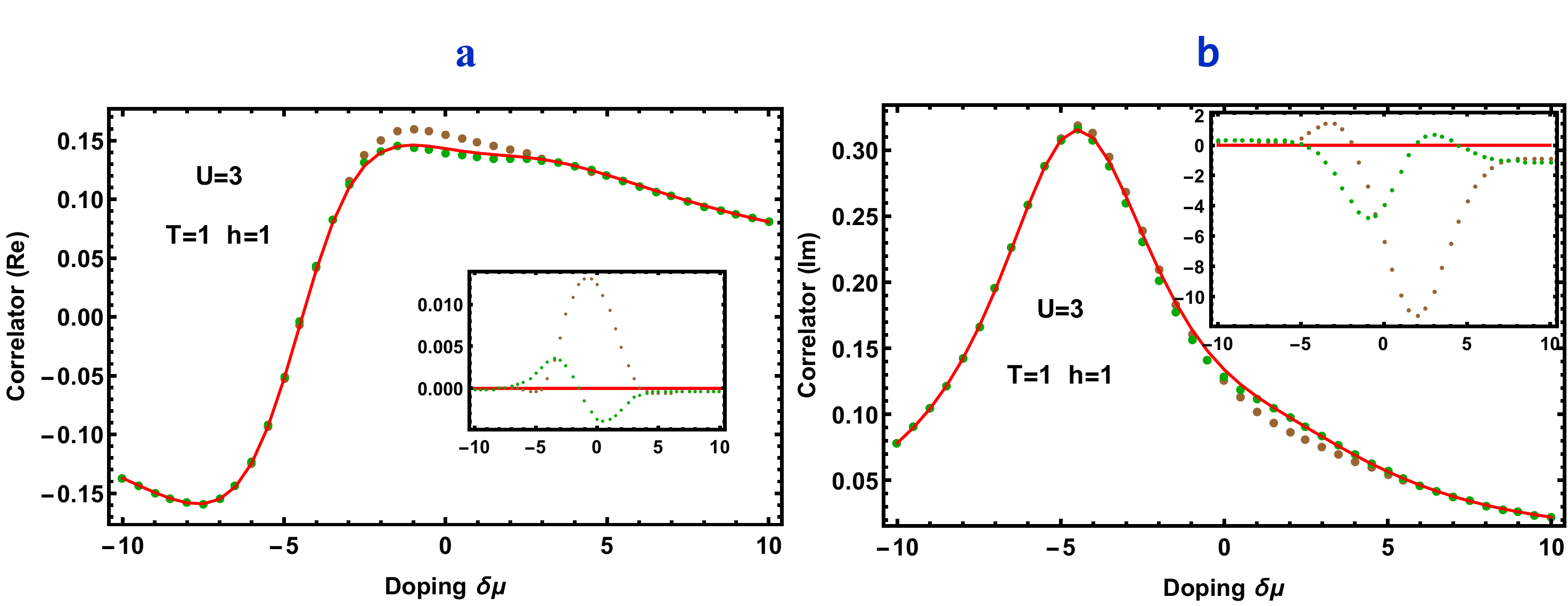}
\end{center}
\par
\vspace{-0.5cm}
\caption{Fig.8. Real (a) and imaginary (b) parts of the Matsubara correlator
for the Hubbard spin chain at intermediate coupling .}
\end{figure}

\begin{figure}[tbp]
\begin{center}
\includegraphics[width=18cm]{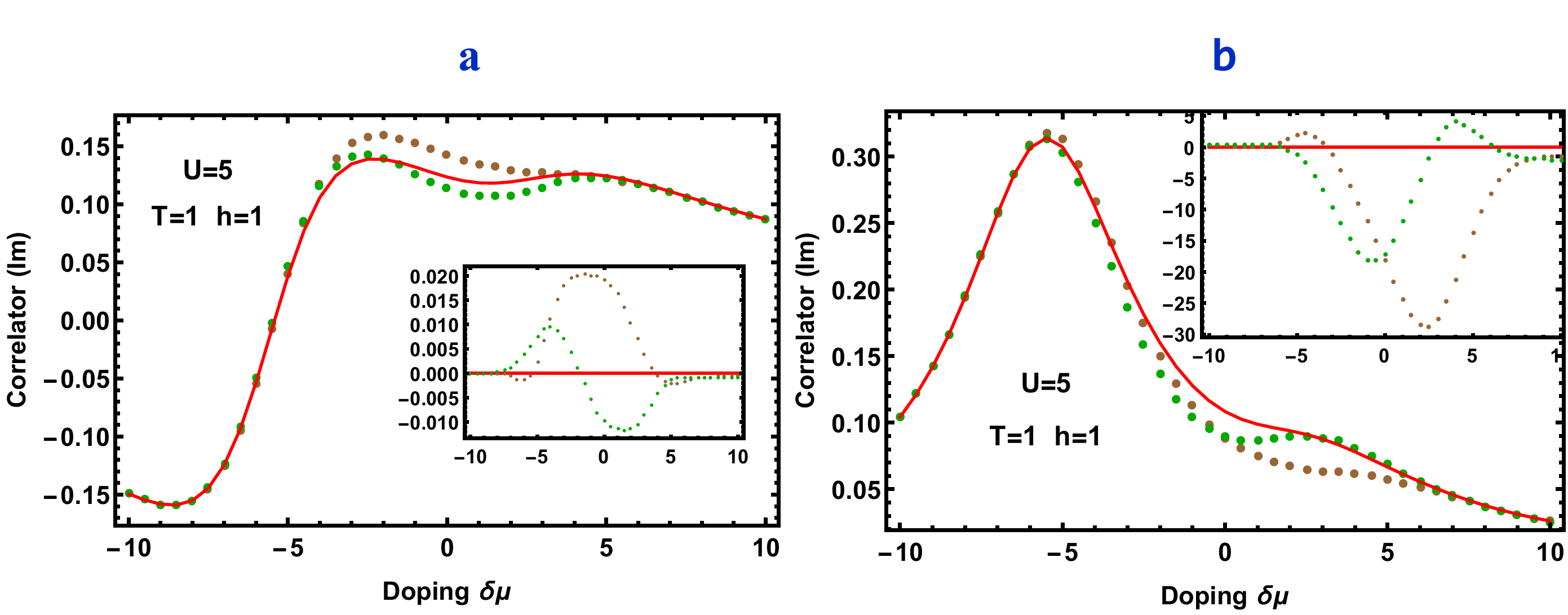}
\end{center}
\par
\vspace{-0.5cm}
\caption{Fig.9. Same as Fig. 8 for a larger value of the coupling.}
\end{figure}

\subsection{Comparison of results with exact diagonalization for one
dimensional Hubbard model}

The one dimensional case is considered for simplicity and availability of
exact results utilizing the exact diagonalization\cite{ED} for reasonably
large values of $N_{s}$. The largest lattice we have used to exactly
calculate Green's function was $N_{s}=6$ (so that the number of fermionic
degrees of freedom is $12$) and avoided using the Lancos algorithms to
diagonalize large matrices, since temperature range $T=0.2-4$ in units of
the hopping parameter $t$ of the Hubbard model is considered. The relatively
high temperature allows lower $N_{t}=512-2048$ to obtain precision of $0.2\%$
for the CCA calculation using the ``naive" discretization of Eq.(\ref%
{Liuaction}). Larger lattices were treated by the exact diagonalization,
however in these calculations typically only spectrum and expectation values
were computed. To compare with CCA the Matsubara time correlators are
required. These are more difficult to compute. The program was written in
Mathematica.

The program for CCA was written in C++ and utilizes parallel computing on a $%
128$ node cluster and large memory of $512Gbyte$. The doping range was $%
10<\delta \mu \equiv \mu -U/2<10$ for the band with $4$ (in units of the
hopping parameter). Temperature and magnetic field were fixed at $T=1$, $h=1$%
. The results for $N_{s}=4$ are presented in Fig. 8 and 9 for intermediate, $%
U=3$ and strong, $U=5$, couplings respectively. As in Figs. 8,9 a show the
real part of the Matsubara correlator, while b - the imaginary part. One
observes a very good agreement not only in perturbative domains for large
absolute value of $\delta \mu $ (far away from half filling. The maximal
deviations are $0.013$, $0.003$ (real part) and $12\%$, $5\%$ (imaginary
part) for CGA and CCA respectively for $U=3$, see insets In Fig. 8. In Fig.
9 the same is given for a strong coupling for $U=5$. The agreement is worse,
by still have maximal deviations are $0.02$, $0.01$ (real part) and $28\%$, $%
17\%$ for CGA and CCA. Generally results are similar to that in $D=0$ for
the relatively low value of $N_{s}=4$.

Till now the application of CCA to a number of solvable field theoretical
models was considered to gauge its precision and complexity. It is not the
purpose of the present paper to apply the method to a realistic material,
however below we estimate the mathematical/computational complexity of such
a calculation.

\section{Discussion and conclusions}

\subsection{Complexity of the CCA computation for a realistic material}

Naive estimate of the computational complexity of covariant third order
approximation is misleading due to the following three observations of the
formalism presented in Section IV.

\textit{1.} Since the odd order fermionic correlators vanish, the only
variational parameter is still the truncated correlator, namely the most
complicated third equation, Eq.(\ref{thirdcon}), is trivially satisfied in
the fermionic model (unlike in bosonic model in the symmetry broken phase
that is indeed too complicated). Moreover the only on shell equation
coincides with HF.

\textit{2}. The chain equations, \ref{chainft}, are all linear. We do not
have a proof, but it seems to be a common feature for both covariant
gaussian and CCA.

\textit{3.} The chain equations, albeit linear have a lot of variables and
one can either apply a solution algorithm or look for a convergent
recursion. At least in Hubbard type models such a recursion exists.

Before performing the estimate of memory and time requirements, let us
review the whole procedure depicted in Fig. 5.

\subsubsection{Downfolding}

Let us assume that one would like to calculate the electron Matsubara
correlator for a certain material. This will describe (after necessary
analytic continuation to physical frequency\cite{Gubernatis} $\omega
\rightarrow -i\omega $) the photo - emission, the STM and X-ray diffraction
techniques and other linear response data. The first step, see Fig. 5, would
be to compute the downfolded action. In includes the usual optimization of
the lattice parameters within DFT performed on commercially available
platforms like VASP. This step is common to a large variety of similar
approximation, so we just refer to available literature. It is used just to
project out ``irrelevant" sectors of the Hilbert space.

The DFT Hamiltonian is not used beyond this and the downfolded models
tunneling amplitudes $T^{AB}$ and (partially screened by high energy degrees
of freedom) Coulomb interactions $V^{AB}$ basis set, while Wannier90 inputs
maximally- localized Wannier functions following the method of Marzari and
Vanderbilt\cite{Vanderbilt}, as is implemented for example in Wannier90\cite%
{Wannier}. Then the following CCA steps, see the flowchart, Fig.5, should be
implemented. To make the discussion more specific, let us estimate the
complexity, and provide numbers using an example of a simple 2D
semiconductor hexagonal 2D boron - nitride, layer hBN. In this case one can
retain only four bands, two for the boron atom and two for the nitrogen, so
that the band index takes $N_{b}=8$ including spin. The Brillouin zone grid
contains $N_{s}^{D}$ for $D=2,N_{s}=8$, while number of Matsubara
frequencies is $N_{t}=64$. The later determines the lattice size for a
periodic boundary conditions and is related to physical frequencies.

\subsubsection{ Nonlinear minimization (Hartree - Fock) equations.}

Number of fermionic variables in Matsubara action is%
\begin{equation}
n=N_{s}^{D}N_{t}N_{f}\text{.}  \label{HFest}
\end{equation}%
Therefore one has to solve $n$ nonlinear equations, Eq.(\ref{chainFT}). This
is typically done very effectively iteratively and DFT software often
provide the result. Of course, as explained in detail in Section IV, HF
approximation is not successful in many instances, but in the present
approach constitutes only the first step. For BN the number of equations is $%
n=32768$ based $N_{s}$, $N_{t}$, $N_{f}$ assumed. Therefore generally there
is no problem with either memory of time of calculation for this simple case.

\subsubsection{Linear chain equations solution.}

The price to pay is that in addition to computing the HF fermionic Green's
function $G_{\omega k}^{trAB}$, one also has to solve an extensive system of
linear equations, the so-called chain equations, either in the configuration
space, Eq.(\ref{Dchain}), or frequency - $k$ - vector space, Eq.(\ref%
{chainFT}). The chain correction is then added to the inverse Green's
function that is inherently charge conserving. The number of ``chains" after
reductions due to translation symmetry is very large. In the absence of
symmetries (like the spin symmetry etc) the number of variables in the chain
equations, Eq.(\ref{chainFT}), is:%
\begin{equation}
n_{ch}=2N_{b}^{3}N_{s}^{2D}N_{t}^{2}  \label{number_chains}
\end{equation}%
The factor $2$ is due to two ``charge" channels, cooperon and diffuson. In
the hBN example it amounts to $n_{ch}=2\times 8^{3}\times 8^{4}\times
64=2.6\times 10^{8}$.

However the matrix is sparse matrix, since there is only one space and time
summation in the chain equation, Eq.(\ref{chainFT}). The density of the
matrix (ratio of nonzero matrix elements of the total $n_{ch}^{2}$) is
\begin{equation}
density=\frac{1}{N_{t}N_{s}^{D}}\text{.}  \label{matdensity}
\end{equation}
This amounts to $2.4\times 10^{-4}$. Of course symmetries sometimes reduce
the number, but in order to solve the equations with ``exact" algorithms like
that in the Intel LAPACK package, large memory and extensive parallel
computing are required unless the matrices of these linear equations are not
very sparse. Planned calculation uses workstation with $2.Tbyte$ active
memory with $128$ Intel $2.8$ $THz$ nodes.

The matrices are expected to be sparse in the configuration space, Eq.(\ref%
{Dchain}), since coefficients contain many fast decreasing Matsubara
correlators. We have not made use of this calculations, preferring exact
solution by LAPACK. However in realistic calculations, one might have to use
properly constructed iteration scheme. To calculate the whole set of
frequencies (required for the analytic continuation) and $k$-vectors the
equations should be solved $n$ times (``spectator" parameter in the CCA
scheme, see Section IV). Other method, like iteration (using fast Fourier
transforms) might be much faster.

Another possibility to optimize the calculation would be to use so called
improved fermionic action\cite{Drut}, so that $N_{t}$ can be reduced. It has
been successfully applied in condensed matter simulations. No results for
improved actions are presented here. Note however that since continuation
(imaginary to real time ) to physical frequencies should be made $N_{t}$
cannot be too small.

\subsection{Conclusions}

To summarize, we have developed a non - perturbative manifestly charge
conserving method, covariant cubic approximation, CCA, determining the
excitation properties of crystalline solids is proposed. Although the basic
band structure of crystalline solids can be theoretically investigated by
the density functional methods, the condensed matter characteristics
dependent on the detailed structure of the electronic matter near the Fermi
level requires more precise treatment of the electrons near the Fermi level.
Like some other methods (versions of GW and various Monte Carlo based
methods) CCA relies on ``downfolding" of the original microscopic model to a
simpler electronic model on the lattice with pairwise interactions. Thus DFT
is used only to deduce the ``downfolded" model on the lattice with pairwise
interacting electrons on a limited set of relevant electronic bands.

It was shown that truncation of the set of Dyson - Schwinger equations for
correlators of the downfolded model of a material lead to a converging
series of approximates. The covariance ensures that all the Ward identities
expressing the charge conservation are obeyed. A large number of solvable
bosonic and fermionic field theoretical models demonstrate that the third
approximant in this series, CCA, is sufficiently precise. Moreover it turns
out that is still calculable by currently available calculational tools. We
focused here on the electron correlator describing single electron (hole)
excitations observed directly by for example the photoemission experiments.

\section*{Acknowledgment}

Authors are very grateful to J. Wang, I. Berenstein, H.C. Kao, B. Shapiro,
T. X. Ma, Y. H. Chiu, G. Leshem for numerous discussions and help in
computations. B. R. were supported by MOST of Taiwan through Contract Grant
104-2112-M-003-012. D. P. L. was supported by National Natural Science
Foundation of China (No. 11674007 and No. 91736208). B.R. is grateful to
School of Physics of Peking University for hospitality.


\begin{thebibliography}{99}
\bibitem{footnote1} For negative $a$ the exact expression takes a slightly
different form. Negative $a$ leads to the field reflection ($\psi
\rightarrow -\psi $) $Z_{2}$ - spontaneously broken solutions in the
classical and simetimes gaussian approximation. Despite the fact that we
know there is no spontaneous symmetry breaking in $d=0$, a rather precise
approximation scheme for a $Z_{2}$ invariant quantity emerges when one of
the two \textquotedblleft would be" symmetry broken solutions, $\varphi =\pm
\sqrt{-a/b}$, is often considered in the intermediate steps. While in a
rather more comprehensive description of covariant gaussian approximation in%
\cite{Wang17} the focus has been also on the symmetry broken phases, in the
present work the emphasis will be on detailed structure of the ``two - body"
correlator in the symmetry preserved phase.
\end{thebibliography}
\end{document}